\documentclass[letterpaper]{aa520} 

\usepackage{psfig}
\usepackage{epsf}
\usepackage{bigdelim}
\usepackage{bigstrut}
\usepackage{multirow}
\usepackage{txfonts}
\usepackage{aalongtable}
\usepackage{natbib}

\def\HI{\ifmmode{\rm HI}\else{H\/{\sc i}}\fi}

\def\magasas{\nobreak\mbox{$\;$mag$\,$arcsec$^{-2}$}} 
\def\msun{\ifmmode{{\mathrm M}_{\odot}}\else{M$_{\odot}$}\fi} 
\def\deg{\hbox{$^\circ$}}
\def\arcmin{\hbox{$^\prime$}}
\def\arcsec{\hbox{$^{\prime\prime}$}}
\def\msunpc2{\ifmmode{{\mathrm M}_{\odot} \, {\mathrm{pc}}^{-2}}\else{M$_{\odot} \, {\mathrm {pc}}^{-2}$}\fi}

\bibpunct{(}{)}{;}{a}{}{,}

\defcitealias{Swaters02}{Paper~I}
\defcitealias{WHISPII}{Paper~II}

\begin{document}

\title{The Westerbork HI survey of spiral and irregular galaxies}
\subtitle{III: HI observations of early-type disk galaxies} 

\author{E.\ Noordermeer\inst{1}\thanks{\mbox{email: edo@astro.rug.nl}} \and
J.M.\ van der Hulst\inst{1} \and 
R.\ Sancisi\inst{1,2} \and R.A.\ Swaters\inst{3} \and
T.S.\ van Albada\inst{1}}

\institute{Kapteyn Astronomical Institute, University of Groningen,
  P.O. Box 800, 9700 AV Groningen, The Netherlands \and
  INAF-Osservatorio Astronomico di Bologna, Via Ranzani 1, 40127
  Bologna, Italy \and Department of Astronomy, University of Maryland, 
  College Park, MD 20742-2421, U.S.A.}   

\date{Received 9-9-9999 / Accepted 9-9-9999}

\abstract{
  We present \HI\ observations of 68 early-type disk galaxies from the
  WHISP survey.   
  They have morphological types between S0 and Sab and absolute
  B-band magnitudes between -14 and -22.
  These galaxies form the massive, high surface-brightness extreme of
  the disk galaxy population, few of which have been imaged in \HI\ 
  before. 
  The \HI\ properties of the galaxies in our sample span a large
  range; the average values of ${\mathrm M}_\HI / {\mathrm L}_B$ and  
  ${\mathrm D}_\HI/{\mathrm D}_{25}$ are comparable to the ones found
  in later-type spirals, but the dispersions around the mean are
  larger.  
  No significant differences are found between the S0/S0a and the
  Sa/Sab galaxies.  
  Our early-type disk galaxies follow the same \HI\ mass-diameter
  relation as later-type spiral galaxies, but their effective \HI\
  surface densities are slightly lower than those found in later-type
  systems.     
  In some galaxies, distinct rings of \HI\ emission coincide with
  regions of enhanced star formation, even though the average gas
  densities are far below the threshold of star formation derived by 
  \citet{Kennicutt89}. 
  Apparently, additional mechanisms, as yet unknown, regulate star
  formation at low surface densities.
  Many of the galaxies in our sample have lopsided gas morphologies;
  in most cases this can be linked to recent or ongoing interactions
  or merger events. Asymmetries are rare in quiescent galaxies.
  Kinematic lopsidedness is rare, both in interacting and isolated
  systems. 
  In the appendix, we present an atlas of the \HI\ observations: for
  all galaxies we show \HI\ surface density maps, global
  profiles, velocity fields and radial surface density profiles. 
}

\maketitle

\section{Introduction}
\label{sec:introduction}
The discovery that \HI\ rotation curves of spiral galaxies remain flat
far outside the regions where the luminous mass is concentrated
\citep{Bosma78}, has led to the now widely accepted view that galaxies
are embedded in large halos of invisible matter \citep{Faber79,
  Bosma81, Albada86}.    
The nature of this dark matter still remains a mystery, and poses one
of the most persistent challenges to present day astronomy.

A systematic study of rotation curves in spiral galaxies, covering a 
large range of luminosities, morphological types and surface
brightnesses, is crucial for a proper understanding of the
distribution of dark matter in galaxies.
Most \HI\ studies in recent years have, however, focused on late-type,
low-luminosity systems \citep{DeBlok96, Swaters02}; \HI\ observations
of high-luminosity and early-type disk galaxies are rare.   
The only study so far aimed at a systematic investigation of \HI\
rotation curves over the full range of morphological types was that
by \citet{Broeils92}. However, in his sample of 23 galaxies, only 
one was of morphological type earlier than Sb and only four had
$V_{\mathrm {max}} > 250 \, {\mathrm {km \, s^{-1}}}$.  
S0 and Sa galaxies were also under-represented in the study by
\citet{Persic96}; their Universal Rotation Curve is based on over
1000 rotation curves of which only 2 are of type Sab or earlier.  

The main reason for the lack of information on the \HI\ properties of
early-type disks is the fact that they generally have a low \HI\
flux \citep{Roberts94}. 
The only large-scale \HI\ survey directed specifically at S0 and Sa
galaxies was carried out by \citet{VanDriel87}, but his study was
severely hampered by the low signal-to-noise ratio of his data.
\citet{Jore97} obtained deep VLA observations for a sample of Sa
galaxies, but focused mostly on asymmetries and evidence for
interactions and mergers \citep{Haynes00}. 

Early-type disk galaxies form the high-mass, high-surface-brightness 
end of the disk galaxy population \citep{Roberts94}. 
While the late-type spiral and irregular galaxies at the other end
of the population are generally believed to be dominated by dark
matter \citep{Carignan88, Broeils92, Swatersthesis}, little yet is
known about the relative importance of dark and luminous matter in
early-type disks.   
Knowledge of their dark matter content is a crucial step towards a
proper understanding of the systematics of dark matter in galaxies.

The work presented here is part of a larger study aimed at
investigating the properties of dark matter in early-type disk
galaxies and its relation with the luminous components.  
A major part of our study consists of mapping the distribution  
and kinematics of neutral hydrogen in a large sample of these
systems. 
As part of the WHISP project, we have observed 68 galaxies with
morphological types between S0 and Sab.   
WHISP \citep[Westerbork \HI\ survey of spiral and irregular
galaxies;][]{Kamphuis96, VanderHulst01} is a survey of galaxies in
the northern sky using the Westerbork Synthesis Radio Telescope
(WSRT), aiming at a systematic investigation of the \HI\ content and 
large-scale kinematics of disk galaxies.  
Observations were carried out between 1993 and 2002, and the survey
now contains data for almost 400 galaxies, covering Hubble types from 
S0 to Im. 
Earlier results from WHISP were presented by \citet[][ hereafter
Paper~I]{Swaters02}, \citet[][ Paper~II]{WHISPII} and
\citet{Garcia-Ruiz02}.    

In this paper, we present the first results of the \HI\ observations
of the early-type disk galaxies. 
We discuss the basic \HI\ properties of these galaxies and we present  
an atlas of \HI\ surface density maps, velocity fields, global
profiles and radial surface density distributions. 
The data presented here will be used in a forthcoming paper to derive  
rotation curves, which will then be combined with optical data to
study the dark matter content and distribution in these systems.  

The structure of this paper is as follows. 
In Sect.~\ref{sec:sample} we describe the selection of the sample
and some basic properties of the selected targets. 
In Sect.~\ref{sec:data}, the observations and data reduction steps 
are discussed.  
In Sect.~\ref{sec:results}, we describe the \HI\ properties of our 
sample galaxies, and compare them to their optical properties.  
Notes on individual galaxies are given in Sect.~\ref{sec:notes}.  
Section~\ref{sec:conclusions} presents a summary of the main
conclusions.
The atlas of the \HI\ observations is presented in the appendix.

\section{Sample selection}
\label{sec:sample}
All galaxies presented here were selected from the WHISP survey
\citep{Kamphuis96, VanderHulst01}.   
The galaxies in WHISP were selected from the Uppsala General Catalogue 
of Galaxies \citep[UGC;][]{UGC}. 
To ensure that the galaxies would be well resolved with the WSRT,
target galaxies were selected on the basis of their position on the
sky ($\delta > 20^\circ$), and angular size (D$_{25} > 1\arcmin$). 
A lower line flux limit, defined as $f = F_{\HI}/W_{20}$ with
$F_{\HI}$ the total flux in ${\mathrm {Jy \, km \, s^{-1}}}$ and
$W_{20}$ the width of the \HI\ profile in ${\mathrm {km \, s^{-1}}}$,
was chosen to ensure that the channel maps have a sufficient 
signal-to-noise  
ratio to produce \HI\ surface density maps and velocity fields. 
For the largest part of the survey, carried out before 1999, this
limit was set at $f>100$~mJy.  
This led to the inclusion of only 11 early-type galaxies, compared
to over 300 intermediate and late-type galaxies. 
In 1999, a more powerful correlator and cooled front-ends on all 
WSRT telescopes improved the sensitivity for 21cm line observations 
by approximately a factor of~3. 
For the observations with the new setup, we lowered the flux limit
to $f>20$~mJy.

From the final WHISP sample, we selected all 68 galaxies with
morphological type between S0 and Sab. Some basic properties of these
galaxies are listed in Table~\ref{table:basicdata}.     

\begin{figure*}[htb]
  \centerline{\psfig{figure=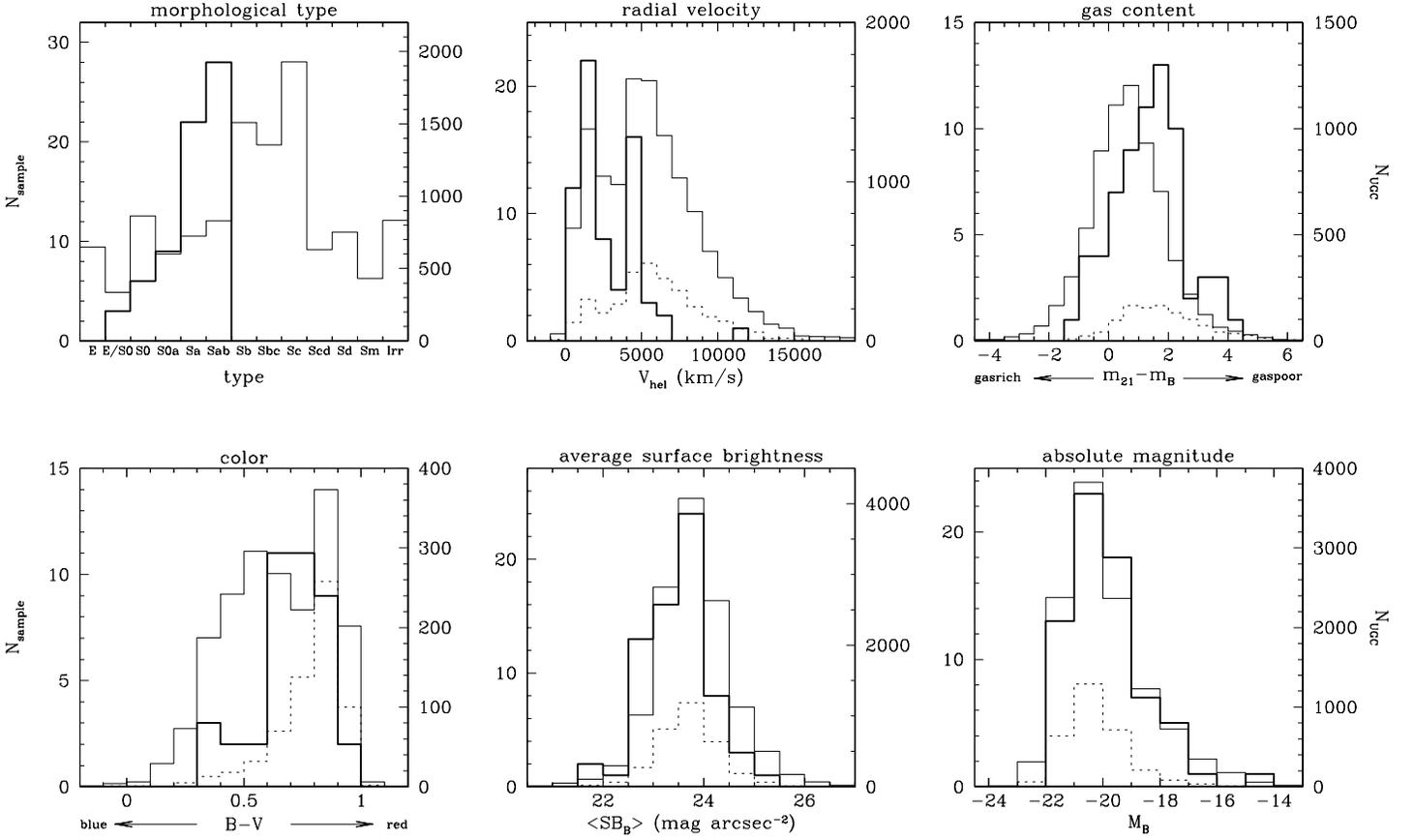,height=4.5in,angle=-90}} 
  \caption{Characteristics of our sample galaxies (bold lines),
  compared with the entire UGC (thin lines) and with all galaxies from  
  the UGC with type between S0 and Sab (dashed lines). The scales on
  the left of each panel give the distribution of the galaxies in the
  sample presented here, the scales on the right apply to the galaxies
  from the UGC.\setcounter{footnote}{0} All data are taken 
  from LEDA\protect\footnotemark, except the heliocentric velocities
  for our sample galaxies, which are taken from this study. Note that
  some parameters (e.g.\ color, gas content) are not available for all 
  galaxies in LEDA; the total number of galaxies included is different
  for each histogram.
  \label{fig:6histograms}}
\end{figure*}

Our sample is not intended to be complete in any sense, but it is
still interesting to compare some general properties of our objects to  
the full sample of galaxies in the UGC. 
In Fig.~\ref{fig:6histograms}, we show in bold lines the
distribution of our sample galaxies over a number of parameters. We
also show the distribution of the entire UGC catalogue (thin lines),
as well as all galaxies from the UGC with morphological type between
S0 and Sab (dashed lines). 
 
The histogram for morphological type simply reflects our
selection criterion for  
early-type disk galaxies. Compared to the entire early-type disk 
subsample from the UGC, we have a fair sample of Sa and Sab galaxies;
S0's are however underrepresented in our sample. This is due to the 
well-known fact that S0 galaxies have generally an even lower \HI\ 
content than Sa's \citep{Roberts94}; even with the current high
sensitivity of the WSRT, only a handful of S0 galaxies can be
observed in single 12h runs. 

Due to our diameter and \HI\ flux limits, we select predominantly 
nearby galaxies, as is visible in the top-middle histogram in 
Fig.~\ref{fig:6histograms}. Most of our sample galaxies have 
\mbox{$V_{\mathrm {hel}} \leq 5000 \, {\mathrm {km \, s^{-1}}}$.}  

Compared to the average early-type disk galaxies in the UGC, our
objects are on average slightly more gas-rich (probed by the
$m_{21}-m_{\mathrm B}$ `color') and marginally bluer. The distribution
of average B-band surface brightness of our galaxies, defined within the
25th magnitude isophote, shows a small excess on the bright side, but 
this is hardly significant. Similarly, we have a small excess of 
low-luminosity systems.

In conclusion, our sample contains a fair number of Sa and Sab
galaxies, whose properties seem to resemble those of average
early-type disk galaxies in the UGC very well.  
Their bluer color and higher gas content may indicate a higher
level of star formation, but the differences for both parameters are
small. 
The fact that the average surface brightness and luminosity
distributions are almost identical to those for the entire subsample
of early-type disks from the UGC, confirms our conjecture that we are
not looking at a special subclass of these systems. The
situation for S0's is less clear, due to the small number contained in
this sample.

\section{Observations and data reduction}
\label{sec:data}
All observations described here were carried out with the Westerbork
Synthesis Radio Telescope (WSRT) in single 12 hour sessions. 
11 galaxies were observed before 1999, using the old front-ends and 
correlator. 
The remaining 57 were observed after the major upgrade of the WSRT,
using new cooled front-ends on all telescopes and a new broadband
correlator.  
A summary of the observational parameters is given in
Table~\ref{table:observations}. 

For most observations, the field of view was centered on the target
galaxy.  
In a few cases, more than one galaxy was observed in one pointing, and
some galaxies lie off-center in the field of view (notably UGC~94,
499, 508, 6621, 8699 and 12815). 
The angular resolution depends on the coverage of the UV-plane
during the observations. Furthermore, as the WSRT is an east-west
array, the resolution in declination depends on the position on the
sky and scales as $1/\sin \delta$. The resulting beam size for each
galaxy is given in Table~\ref{table:observations}. 

Observations were usually done with 128 channels; the total bandwidth
was chosen on the basis of the \HI\ line width $W_{20}$, such that we  
had optimal wavelength resolution while still having a reasonable
baseline to construct continuum maps. The resulting velocity
resolution, before Hanning smoothing, is thus either $\sim$ 2.5, 5, 10
or 20 km s$^{-1}$, with the exact value depending on the central wavelength
of the band (see Table~\ref{table:observations}). 

The data reduction consisted of several stages; each will be described  
below. 
\footnotetext{LEDA: Lyon Extragalactic Database,
  http://leda.univ-lyon1.fr/}

\subsection{standard WHISP pipeline}
\label{subsec:WHISP pipeline}
The initial reduction steps were carried out following the standard 
WHISP data reduction pipeline. The raw UV data were inspected and 
calibrated using the NEWSTAR software package. Bad data-points were 
flagged interactively. The UV data were then Fourier transformed to
the image plane, and antenna patterns were calculated.   
For each set of observations, data cubes were created at 3 different 
resolutions and corresponding noise levels. One is at full resolution 
using all available UV data-points; the resulting beam size is given
in Table~\ref{table:observations}. 
The other two were produced by down-weighting progressively more long
baselines from the UV data.  
While doing so, the spatial resolution was decreased to respectively
$\approx 30\arcsec$ and $\approx 60\arcsec$, but the signal-to-noise
level for extended emission was strongly enhanced. 

All further data reduction was done with GIPSY 
\citep[Groningen Image Processing System;][
  http://www.astro.rug.nl/$\sim$gipsy/]{Vogelaar01}.    
For all data cubes, the continuum was subtracted by fitting, at each
line of sight, a first order polynomial to the channels without \HI\
emission.
The cube at $60\arcsec$ resolution was then Hanning smoothed in
velocity and masks were created by hand to identify the regions
containing \HI\ emission. 
The data cubes were then CLEANed \citep{Schwartz78}, using the masks to
define the search areas.  
CLEANing was iterated down to 0.5 times the rms in each channel map.
The rms noise in the cleaned, full resolution channel maps,
$\sigma_{\mathrm{ch}}$, is given in Table~\ref{table:observations}.

In more than 50\% of the cases, the data cubes at full resolution are 
of sufficient quality to derive useful \HI\ maps and velocity fields.  
However, in cases where the signal is very weak, the signal-to-noise
ratio in the full resolution data may be too low to extract the
emission line profiles.  
In those cases we used the smoothed data in the subsequent analysis. 
Additionally, the smoothed data cubes are sometimes used to study
the distribution and kinematics of the gas in the faint outer regions.  
In the figures in the atlas we indicate which data cube was used for
each galaxy. 

\subsection{defining masks}
\label{subsec:masks}
Before proceeding to the next steps, we defined first the regions that
contain emission; parts of the data cubes that contain only noise were
masked out, such that as little noise as possible entered the global
profiles and surface density maps.  
 
Creating the masks is not straightforward and requires special care. 
Applying a simple sigma-clipping criterion on the channel maps is not
sufficient, as we may miss extended low-level emission.  
Therefore, we first smoothed the channel maps: the channel maps at
full, 30 and $60\arcsec$ resolution were smoothed to 30, 60 and
$120\arcsec$ respectively. 
Subsequently the data cubes were Hanning smoothed in velocity. 
The signal-to-noise ratio for extended emission in the
resulting smoothed data cubes is much higher and low-level emission is
now easily detected.   

The Hanning-smoothed data cubes at 30, 60 and $120\arcsec$ were
then used to create masks for the original channel maps at full, 30 and
$60\arcsec$ resolution respectively; the masks were created in 2
steps. 
First we selected all pixels in the smoothed channel maps that have
intensity $> 2 \sigma$.  
In this way, we did not only select real emission from the galaxy, but
also many noise peaks in the data.  
In the second step we selected by hand the emission which we deemed to
be real.  
Emission was defined to be real if it is spatially extended (i.e. $\ga
2$ beam areas) and present in more than 2 adjacent channels {\em or}
when it is very bright ($\ga 4 \sigma$). 
Small regions that are only slightly brighter than $2 \sigma$ are
defined to be noise and were rejected. 
This second step is by definition subjective. 
Some peaks we judged to be noise may actually be real and vice versa. 
However, since we are working on extensively smoothed data, the real
emission is more easily distinguished from noise, and the effect is
small.  

The resulting clipped channel maps were then used as masks for the
original data cubes.

\subsection{global profiles and total \HI\ masses}
\label{subsec:globprofs}
Global profiles were derived from the data cubes at $60\arcsec$
resolution.     
In these data cubes, we are most sensitive to extended \HI\ emission,
and the flux enclosed by the masks is maximal.  
The global profiles were constructed by adding up all flux in each
masked channel map and correcting for primary beam attenuation.  
Care was taken not to include flux from companion galaxies.  

The error in each point of the profile is related to the size of the 
masked region in the channel map. 
Due to the non-uniform sampling of the UV-plane by the WSRT, the noise
in the channel maps is not spatially independent and the error does
{\em not} simply increase as  the square root of the mask area. 
Instead, we established an empirical relation between mask size and
error in the flux and used the result to estimate the errors on our
global profiles. 

The global profiles are shown in the top right panels in the atlas
(Appendix~\ref{app:atlas}).  

Line widths were determined at the 20\% and 50\% levels. 
If a profile was double peaked, the peaks on both sides were used 
separately to determine the 20\% and 50\% levels, otherwise the 
overall peak flux was used. 
The profile widths $W_{20}$ and $W_{50}$ were defined as the
difference between the velocities at the appropriate level on each
side of the profile. 
We followed \citet{Verheijen01b} to correct the observed line widths
for instrumental broadening, assuming an internal velocity dispersion
of the gas of 10~km~s$^{-1}$:     
\begin{equation}
 \begin{array}{rcl}
  W_{20}^c & = & W_{20} - 35.8\left[\sqrt{1 + \left(\frac{R}{23.5}\right)^2} - 1\right] \\[0.4cm]
  W_{50}^c & = & W_{50} - 23.5\left[\sqrt{1 + \left(\frac{R}{23.5}\right)^2} - 1\right],
 \end{array}
\end{equation}
with $R$ the instrumental velocity resolution in km~s$^{-1}$.

The systemic velocities $V_{\mathrm {sys}}$ were defined as the
average of the velocities at the profile edges at the 20\% and 50\%
level.  
Distances $D$ were derived using a Hubble constant of
75~km~s$^{-1}$~Mpc$^{-1}$, correcting the systemic velocities for
Virgo-centric inflow using the values given by LEDA. 
The profile widths, systemic velocities and distances are given in
Table~\ref{table:data}. 

The total \HI\ mass was derived from the total flux, i.e.\ the
integral of the global profile just derived, and is given by: 
\begin{equation}
 M_{\HI} = 236 \cdot D^{2} \cdot \! \int F {\mathrm {dv}},
\end{equation}
where $M_{\HI}$ is in solar masses, $D$ is the distance in Mpc, $F$ is 
the primary beam corrected flux in mJy, and the integral is over the
total bandwidth. 
The derived total fluxes and \HI\ masses are given in 
Table~\ref{table:data} as well. 

\subsection{integrated \HI\ maps}
\label{subsec:HImaps}
\HI\ maps were created by integrating the masked data cubes along 
the velocity direction and correcting the result for primary beam 
attenuation. 
The noise in these maps is not constant: the galaxies presented here
are generally characterized by very large velocity gradients in the
center and slowly varying velocities in the outer parts, so that we
are summing over a variable number of channel maps at different 
positions. 
For each map, we estimated an average noise level,
$\sigma_{\mathrm{map}}$, following the prescriptions in
\citet{Verheijen01b}.   
Noise fields were constructed based on the number of channel maps
contributing to the integrated \HI\ maps at each position and the rms
noise in the individual channel maps. 
Using these fields, we selected all points in our \HI\ maps with
signal-to-noise ratio between 2.75 and 3.25.  
The average value of these points was then divided by 3 to obtain the 
$\sigma_{\mathrm{map}}$ value. 
It is important to note that this is an {\em average} noise level;
low-level emission which is only present in a few adjacent channels may
be significant, even if it is weaker than the
$2\sigma_{\mathrm{map}}$-level of the outer contour in the figures. 

The \HI\ column density maps are shown in the left hand panels in
the atlas. In Table~\ref{table:observations}, we give the 
corresponding values of $\sigma_{\mathrm{map}}$ in atoms~cm$^{-2}$.

\subsection{velocity fields}
\label{subsec:vfields}
The two methods most commonly used to derive velocity fields from \HI\
data cubes are to calculate the intensity weighted mean (IWM) velocity
of the line profiles or to fit Gaussians to them.
For the data presented here, neither of these methods proved to be
adequate in recovering the true radial velocity of the gas. 
Particularly in the central regions of the early-type galaxies studied  
here, large velocity gradients are present and the line profiles
are seriously skewed by beam-smearing. 
In those cases, both the IWM velocity and the central velocity of the
fitted Gaussians can be offset significantly from the true gas
velocity at the projected radius \citepalias[e.g.][]{Swaters02}. 
Here, we derive velocity fields by fitting a Gauss-Hermite polynomial
that includes an h3-term for skewness to the line profiles
\citep{VanderMarel93}. 
The resulting velocities from this method lie usually closer to the
peak in the profile than simple IWM or Gauss fits, especially in the
central regions.

We created velocity fields in two steps. 
In the first step, we fitted the Gauss-Hermite function to the profiles
in the masked data cubes. 
However, since the masked data cubes contain no noise information
outside the emission lines, it is difficult to determine the
significance and reliability of the fitted profiles. 
To overcome these difficulties, we used the fitted parameters from the 
first step as initial estimates for a second fit to the
Hanning-smoothed, {\em unmasked} line profiles.  
For these latter fits we could develop strict criteria to determine
whether or not a profile represents true emission: only those fits
were accepted that have {\it 1)} a line strength greater than 2.5
times the rms noise $\sigma_{\mathrm{h}}$ in the Hanning-smoothed
channel maps and {\it 2)} a velocity within the range defined by the
global profile.  
These two constraints were usually sufficient to exclude erroneous
results.  
As an extra sanity check, we excluded fits with {\it 3)} a profile
width smaller than 4 or larger than 200 km~s$^{-1}$. 
This criterion was, however, rarely met.  

It is important to note here that the sigma-clipping criterion in {\it
1)} above was performed using the rms in the Hanning-smoothed channel
maps directly. 
This rms is not the same as the $\sigma_{\mathrm{map}}$ level in the 
\HI\ surface density maps, which was determined empirically (see
previous section). 
Thus, it is possible that points which fall outside the
$2\sigma_{\mathrm{map}}$ contour in the \HI\ density map, have a
fitted line strength greater than $2.5 \, \sigma_{\mathrm{h}}$ and vice 
versa.  

The final velocity fields are shown in the top middle panels in
the atlas. In the bottom middle panels, we show the fitted velocities
over-plotted on position-velocity slices along the major axes
through the data cubes. 
From these figures, it is clear that the inclusion of an h3-term
for skewness in the fitting procedure worked well in recovering the
true velocity of the gas for most galaxies (e.g.\ UGC~89, 6786). 
In some cases, however, in particular in edge-on or poorly resolved
galaxies, projection effects lead to such complicated line-profiles
that even skewed Gaussians fail to recover the projected rotational
velocity of the gas.  
In our sample, the velocity fields of UGC~1310, 2045, 5906, 5960,
6001, 6742, 7704, 8271 and 12713 suffer seriously from this effect and
can therefore not be used to derive rotation curves; for UGC~2183,
3354, 4605 and 11670, only  the inner regions of the velocity fields
are affected.

\subsection{radial profiles, \HI\ diameters and surface densities}    
\label{subsec:radprofs}
The integrated \HI\ maps created in Sect.~\ref{subsec:HImaps} were  
used to derive radial profiles of the \HI\ surface density. The 
intensities were azimuthally averaged in concentric elliptical
annuli. Pixels without measured signal in the total \HI\ map were 
counted as zero. The azimuthal averaging was also done separately for
the approaching and receding half of the galaxy to obtain a crude
estimate of the level of asymmetries present in the \HI\ map. 

The orientation of the annuli was determined with different methods,
depending on the galaxy. 
For a number of galaxies, tilted ring fits to the velocity field were
available.  
The details of these fits will be described in a forthcoming paper
dealing with the rotation curves of our galaxies. 
If no tilted ring analysis was available (for example if the galaxy
is too poorly resolved, or if the \HI\ gas does not show signs of
circular rotation), we used the ellipticity and position angle from
LEDA.  
However, in some cases LEDA does not give a position angle or is
clearly inconsistent with our data.  
Especially for near face-on galaxies, LEDA's position angles
-- if present -- can be far off from the angles we infer from the
velocity field. 
In those cases, we did not use LEDA's position angle, but rather
determined it by eye from our \HI\ maps and velocity fields.  
  
The measured intensities were corrected to face-on surface densities, 
and converted to units of \msunpc2. The resulting 
radial profiles are shown in the bottom right panels in the atlas.  

For edge-on galaxies, this method does not work, as emission from
different radii is superposed onto the same position on the sky. For
these cases, we used the iterative Lucy deconvolution method developed
by \citet{Warmels88}.  
The \HI\ surface density maps were integrated parallel to the minor axis to
get \HI\ strip integrals. 
These were then converted to surface density profiles using the
iterative deconvolution scheme from \citet{Lucy74}, under the
assumption of axisymmetry for the gas distribution.   

From the radial profiles, \HI\ radii were determined. The \HI\
radius ${\mathrm R}_\HI$ is defined as the radius where the face-on
corrected surface density drops below a value of 1~\msunpc2
(equivalent to $1.25 \cdot 10^{20}$~atoms~cm$^{-2}$).  
The derived radii are given in Table~\ref{table:data}.

Finally, we derived the average \HI\ surface density within the
\HI\ radius \mbox{($<\!\Sigma_\HI\!>_{{\mathrm R}_\HI}$)}, as
well as within the optical radius \mbox{($<\!\Sigma_\HI\!>_{{\mathrm 
R}_{25}}$)}. The optical radii were derived from the absorption
corrected 25th B-band\magasas\ diameters, taken from LEDA. The
average \HI\ surface densities are also listed in
Table~\ref{table:data}.

\section{\HI\ properties of early-type disk galaxies}
\label{sec:results}
As mentioned in the introduction, the data presented here are intended
for a study of rotation curves and dark matter in early-type disk
galaxies.  
In this paper, however, we focus on the global \HI\ properties of our
sample galaxies, and the relation with their optical characteristics;
several aspects of this analysis are discussed below.

\subsection{\HI\ content}
\label{subsec:HImass}
\begin{figure}[tb] 
  \centerline{\psfig{figure=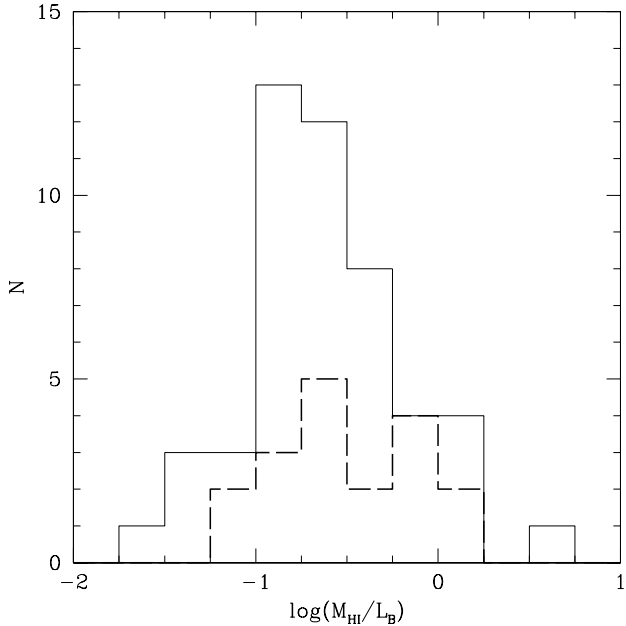,height=3.25in}}
  \caption{The distribution of ${\mathrm M}_\HI / {\mathrm L}_B$. The
  solid line indicates all galaxies of type Sa and Sab, the dashed
  line indicates the S0's.}  
  \label{fig:HImasstolighthisto}
\end{figure}

In Fig.~\ref{fig:HImasstolighthisto}, we show the distribution of
${\mathrm M}_\HI / {\mathrm L}_B$. For ${\mathrm L}_B$, we used the
absolute B-band magnitudes from LEDA (column (6) in
Table~\ref{table:basicdata}) and a solar absolute magnitude in the
B-band of 5.47 \citep{Cox00}. 
For comparison, we have split the sample in S0/S0a's and Sa/Sab's.   

The average values of $\log ({\mathrm M}_\HI / {\mathrm L}_B)$
for Sa/Sab's and S0/S0a's is $-0.62 \pm 0.44$ and $-0.50 \pm 0.40$ in
solar units respectively, where the errors give the standard
deviations of the respective distributions.  
These values are higher than the average values for early-type disks 
found by \citet{Roberts94}, confirming that we have predominantly
selected gas-rich galaxies in our sample (cf.\ Sect.~\ref{sec:sample}
and Fig.~\ref{fig:6histograms}).   
There is, however, a large variation in the \HI\ content of our
galaxies; the most gas-rich galaxies contain about 100 times more gas,
compared to their stellar luminosity, than the most gas-poor
systems. This range is larger than seen in most late-type galaxies,
which span at most only one order of magnitude in gas content
(\citealt{Roberts94}; \citealt{Broeils97}; \citetalias{Swaters02}).  

\begin{figure}[tb] 
  \centerline{\psfig{figure=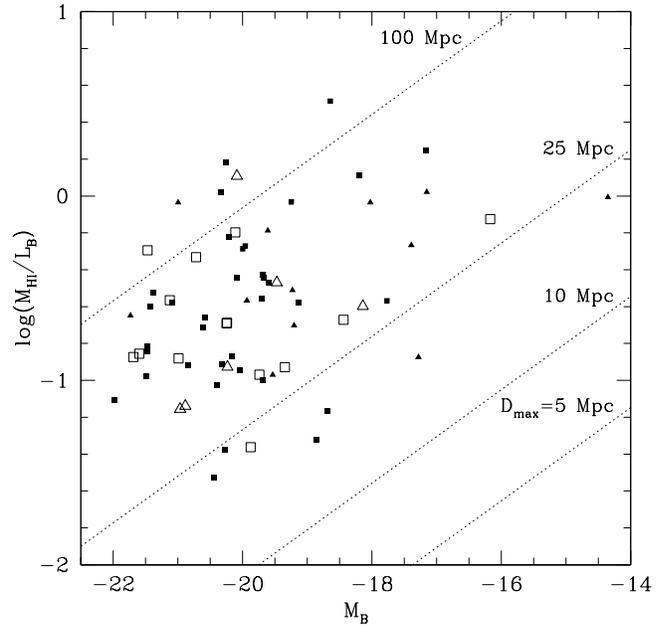,height=3.25in}}
  \caption{${\mathrm M}_\HI / {\mathrm L}_B$ versus absolute B-band
  magnitude. Sa/Sab galaxies are indicated by squares, S0/S0a's by
  triangles. The filled symbols indicate galaxies whose morphology is
  not or only mildly asymmetric, open symbols denote moderately or
  strongly lopsided galaxies. The dashed lines indicate the maximum
  distance at which a galaxy would still be included in our sample
  (see text).\label{fig:HImasstolight_vs_lum}}
\end{figure}

The large spread in \HI\ content is also apparent in
Fig.~\ref{fig:HImasstolight_vs_lum}, where we plot the \HI\
mass-to-light ratio versus B-band luminosity. 
In later-type galaxies, a general trend exists, such that the more
luminous galaxies contain relatively less \HI\
(\citealt{Verheijen01b}; \citetalias{Swaters02}). 
In our sample, this correlation is virtually absent. 
There seems to be a lack of low-luminosity galaxies (${\mathrm {M_B}}
\ga -18.5$) with low relative gas content, but this is probably a
selection effect.   
The dotted lines in Fig.~\ref{fig:HImasstolight_vs_lum} give the
maximum distance ${\mathrm {D_{max}}}$ out to which galaxies would be
included in our sample. 
${\mathrm {D_{max}}}$ is defined as the distance where the flux
density of a galaxy, given by $f = F_{\HI}/W$, drops below 20~mJy. 
$W$, the profile width, is estimated from the absolute magnitude using 
the Tully-Fisher relation found by \citet{Verheijen01a} and assuming 
an average inclination of $60 \deg$.  
It is obvious that high-luminosity galaxies with a low gas content can
be observed out to much larger distances than low-luminosity systems.  
Low-luminosity (${\mathrm {M_B}} \ga -18.5$) galaxies with low
relative gas content ($\log({\mathrm M}_\HI / {\mathrm L}_B)\la -1$)
may be as common as their high-luminosity counterparts, the lack of
low-luminosity, low ${\mathrm M}_\HI / {\mathrm L}_B$ galaxies simply
being the result of the smaller volume that our survey covers for
these systems.    

The large spread in gas content could be explained if a substantial
fraction of our early-type disk galaxies have recently accreted gas
from small dwarf galaxies. 
The spread in gas content would then reflect the accretion history of
these galaxies.  
Indeed, many of our galaxies show signs of recent interaction or
accretion (see Table~\ref{table:data}), but the degree of asymmetry in
the gas distribution does not correlate with the gas fraction, as can
be seen in Fig.~\ref{fig:HImasstolight_vs_lum}.  
It seems unlikely that recent accretion or merger events alone can
explain the observed spread  in gas content of these galaxies. 

Another possible explanation for the large spread lies in the fact
that \HI\ content is a property of the disk, whereas our galaxies
contain an additional bulge component. 
If indeed bulges contribute to the optical luminosity but are not
related to the \HI\ content, variations in bulge-to-disk luminosity
ratios would lead to additional scatter in plots like
Figs.~\ref{fig:HImasstolighthisto} and~\ref{fig:HImasstolight_vs_lum}. 
We have checked this hypothesis for a number of galaxies in our sample 
for which we have accurate optical photometry and bulge-disk
decompositions at our disposal, but found that the scatter in gas
content did not decrease significantly when, instead of the total
luminosity, we only considered the luminosity of the disk component.  

Thus, the \HI\ content in these early-type disk galaxies seems to be 
rather independent of any basic optical property.
In Sect.~\ref{subsec:HIsurfbright}, we will show that the \HI\
content is correlated with the relative extent of the gas disks, in
the sense that the gas disks of gas-rich galaxies are more extended,
but of similar surface density, than those of gas-poor systems. 
Although this implies that the large spread in gas content is, at
least partly, related to the large range in the relative sizes of the
gas disks (cf.\ Sect.~\ref{subsec:HIradii}), it still offers no
explanation as to why some galaxies have larger gas disks, and
correspondingly larger \HI\ masses, than others. 
The origin of the large spread in gas content in these systems remains
therefore puzzling.   

\subsection{\HI\ diameters}
\label{subsec:HIradii}
\begin{figure}[tb] 
  \centerline{\psfig{figure=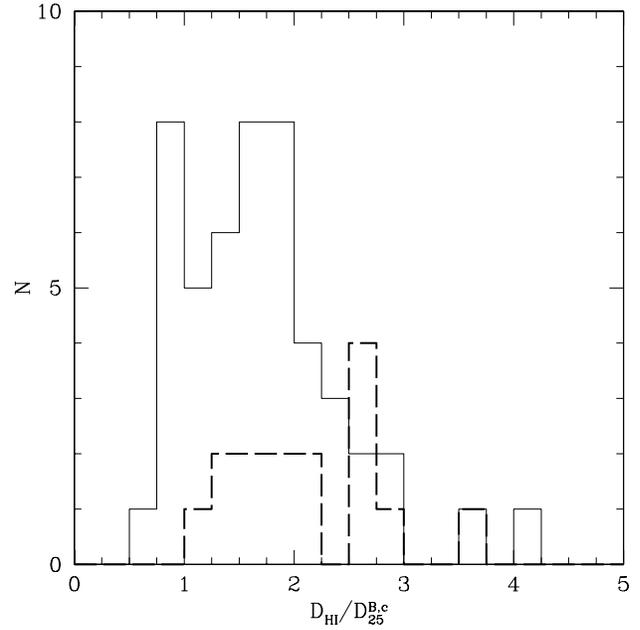,height=3.25in}}
  \caption{The distribution of ${\mathrm D}_\HI/{\mathrm
  D}_{25}^{B,c}$. The solid line indicates all galaxies with type Sa
  and Sab, the dashed line indicates the S0's.}  
  \label{fig:diamhisto}
\end{figure}

In Fig.~\ref{fig:diamhisto}, we show the distribution of 
${\mathrm D}_\HI/{\mathrm D}_{25}^{B,c}$, with ${\mathrm D}_\HI$
the diameters of the \HI\ disks, measured at the 1~\msunpc2 level. 
${\mathrm D}_{25}^{B,c}$ are the optical diameters, measured at the
absorption corrected 25th B-band\magasas\ level, taken from LEDA.   
The average ratio of \HI\ to optical diameter for the 49 Sa/Sab
galaxies in our sample is $1.72 \pm 0.70$; as above, the error gives
the standard deviation of the distribution. This average value is
identical, but with larger spread, to the value found by
\citet{Broeils97} for a sample of 108 spiral galaxies, mostly of
intermediate and late type. 

9 Sa/Sab galaxies (18\%) in our sample have \mbox{${\mathrm D}_\HI <
{\mathrm D}_{25}^{B,c}$.}  
Most of these cases show clear signs of interaction (e.g.\ UGC~4862,
5559, 6621) and it seems likely that these galaxies have lost part of
their gas disks in the encounters.
This would be consistent with the results of \citet{Cayatte94}, who
showed that galaxies in the center of the Virgo cluster have smaller
\HI\ disks than those in the field and argued that this is caused by
ram-pressure stripping and interactions between the cluster members. 
However, some galaxies with small gas disks in our sample seem
undisturbed (e.g.\ UGC~7489 or 11914); the origin of the small extent of
their \HI\ disks is unclear. 

The average ratio of \HI\ to optical diameter for the S0/S0a's in
our sample is $2.11 \pm 0.70$. 
This is slightly higher than for the Sa/Sab's, but the number of
objects is too small to be able to say whether this is significant.  
Note that three S0/S0a galaxies in our sample (UGC~2154, 3426 and
4637) have an azimuthally averaged \HI\ surface density less than
1~\msunpc2\ everywhere, such that their \HI\ radius is not defined;
the gas in these three galaxies is clearly distorted and probably
originates from a recent merger or interaction event. 

\begin{figure}[tb] 
  \centerline{\psfig{figure=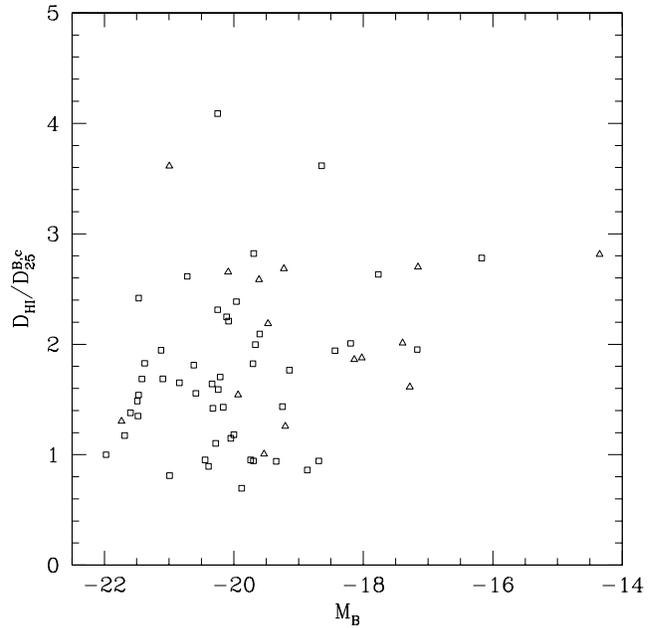,height=3.25in}} 
  \caption{Ratio of \HI\ to optical diameter ${\mathrm D}_\HI/{\mathrm 
  D}_{25}^{B,c}$ versus absolute B-band magnitude. Sa/Sab galaxies are
  indicated by squares, S0/S0a's by triangles.}  
  \label{fig:diamrat_vs_Mabs}
\end{figure}

The ratio ${\mathrm D}_\HI/{\mathrm D}_{25}^{B,c}$ between \HI\ and
optical diameters shows no clear correlation with optical luminosity
(Fig.~\ref{fig:diamrat_vs_Mabs}). 
The lack of low-luminosity galaxies with small \HI\ disks is probably
the result of the same selection effects as described in the previous
section.   

\subsection{\HI\ surface brightness}
\label{subsec:HIsurfbright}
Previous studies have revealed a tight relation between the total \HI\ 
mass and \HI\ diameter of a galaxy, with ${\mathrm M}_\HI \propto
({\mathrm D}_\HI)^\alpha$, \mbox{$\alpha \approx 1.9$}
(\citealt{Broeils97}; \citealt{Verheijen01b};
\citetalias{Swaters02}).
The small scatter around this relation, and the fact that the slope is
so close to 2, imply that the average \HI\ surface density within the
1~\msunpc2\ isophote is almost universal from galaxy to galaxy. 
The correlation of \HI\ mass with optical diameter is well defined
too, but with larger scatter.  

\begin{figure}[tb] 
  \centerline{\psfig{figure=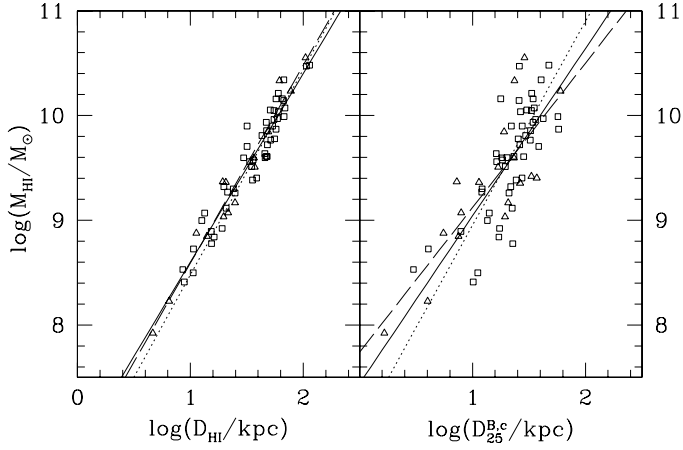,height=2.3in}}
  \caption{Total \HI\ mass versus \HI\ {\em (left)} and optical
  {\em (right)} diameter. Sa/Sab galaxies are indicated by squares, 
  S0/S0a's by triangles; the solid and long-dashed lines give the
  corresponding fits. The dotted lines give the relations found by
  \citet{Broeils97}.}  
  \label{fig:HImass_vs_diam}
\end{figure}

These relations are also present in the early-type galaxies studied
here, as can be seen in Fig.~\ref{fig:HImass_vs_diam}. 
The relation between ${\mathrm M}_\HI$ and ${\mathrm D}_\HI$ is
indistinguishable from the one found in previous studies (the dotted 
line shows the relation from \citet{Broeils97}). We find: 
\begin{equation}
 \begin{array}{c@{\hspace{0.7cm}}l} 
  \log ({\mathrm M}_\HI / \msun)  =  1.80 \log ({\mathrm D}_\HI /
  {\mathrm{kpc}}) + 6.80 &
  {\mathrm{(Sa/Sab)}} \\ 
   \log ({\mathrm M}_\HI / \msun) =  1.89 \log ({\mathrm D}_\HI /
  {\mathrm{kpc}}) + 6.70 &
  {\mathrm{(S0/S0a)}}.
 \end{array}
\end{equation}
The scatter around each of the fits is 0.15 and 0.14 dex respectively,
comparable to the dispersion found by \citet{Broeils97}.  

We find a slightly shallower relation between \HI\ mass and optical
diameter:
\begin{equation}
 \begin{array}{c@{\hspace{0.7cm}}l}
  \log ({\mathrm M}_\HI / \msun) =  1.60 \log ({\mathrm D}_{25}^{B,c} /
  {\mathrm{kpc}}) + 7.44 &
  {\mathrm{(Sa/Sab)}} \\[0.1cm]
  \log ({\mathrm M}_\HI / \msun) =  1.38 \log ({\mathrm D}_{25}^{B,c}
  / {\mathrm{kpc}}) + 7.74 &
  {\mathrm{(S0/S0a)}}.
 \end{array}
\end{equation}
The shallow slopes in these relations could again be the result of a
selection bias in our survey. 
Small, low \HI-mass galaxies are unlikely to be included in our
sample, which explains why the galaxies in our sample with ${\mathrm 
D}_{25}^{B,c} < 10 \, {\mathrm {kpc}}$ lie systematically above the
relation found by \citet{Broeils97}. 
Given our selection criteria, we see no reason to conclude that
early-type disk galaxies follow a different ${\mathrm M}_\HI -
{\mathrm D}_{25}^{B,c}$ relation than later-type spirals.  
Note however that the scatter, 0.34 and 0.38 dex for the Sa/Sab and
S0/S0a samples respectively, is larger than the one found by
\citet{Broeils97}.  

\begin{figure}[tb] 
  \centerline{\psfig{figure=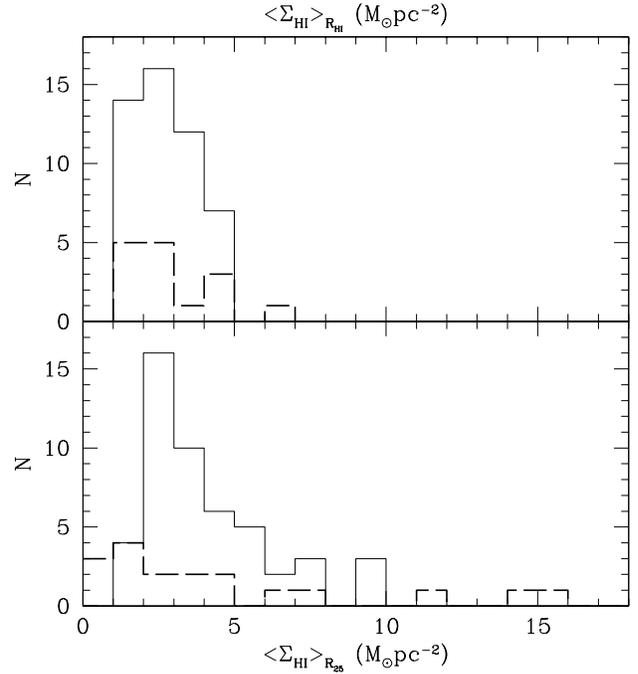,height=3.5in}}
  \caption{The distribution of average \HI\ surface density within
  the 1~\msunpc2\ isophote ({\em top}) and within the optical diameter 
  ({\em bottom}). The solid line indicates all galaxies with type Sa
  and Sab, the dashed line indicates the S0's.}  
  \label{fig:HI_Sig_histos}
\end{figure}

In Fig.~\ref{fig:HI_Sig_histos}, we show the distributions of average
\HI\ surface density within the \HI\ and optical diameters. 
The small scatter in the  ${\mathrm M}_\HI - {\mathrm D}_{\HI}$
relation translates into a narrow distribution in
\mbox{$<\!\Sigma_\HI\!>_{{\mathrm R}_\HI}$}.  
The average values of \mbox{$<\!\Sigma_\HI\!>_{{\mathrm R}_{\HI}}$} 
are $2.8 \pm 0.8$ and $2.9 \pm 1.4\, \msunpc2$ for Sa/Sab 
and S0/S0a galaxies respectively; the errors give the standard 
deviations of the distributions.
Thus, early-type disk galaxies have on average somewhat lower \HI\
surface densities than later-type spirals, but with similar scatter
\citep{Broeils97}.   
The dispersion in the distribution of
\mbox{$<\!\Sigma_\HI\!>_{{\mathrm R}_{25}}$} is much larger, but the 
average values are again lower than in later-type spirals
\citep[cf.][]{Roberts94,Broeils97}.   
 
\begin{figure}[tb] 
  \centerline{\psfig{figure=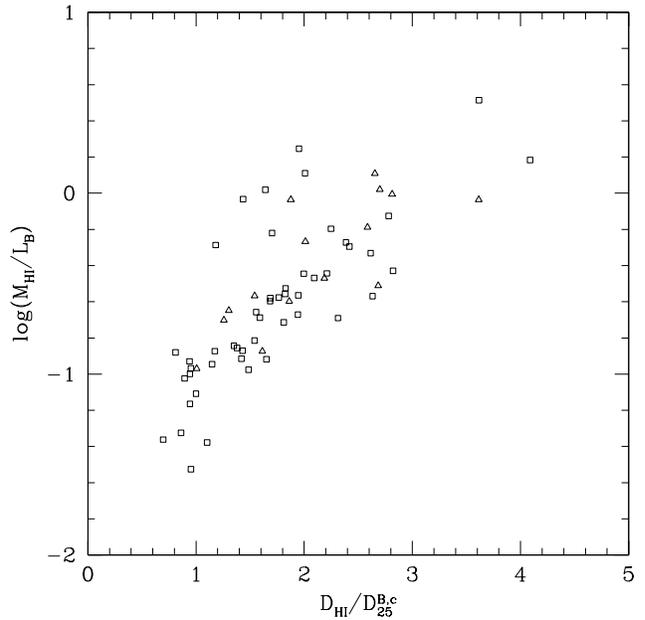,height=3.2in}}
  \caption{Relative gas content, ${\mathrm M}_\HI / {\mathrm L}_B$,
  versus relative gas diameter, ${\mathrm D}_\HI/{\mathrm
  D}_{25}^{B,c}$. Sa/Sab galaxies are indicated by squares, S0/S0a's
  by triangles.}    
  \label{fig:HI_M_L_vs_Drat}
\end{figure}

As a last illustration that the average \HI\ surface density shows
little variation from galaxy to galaxy, we show in
Fig.~\ref{fig:HI_M_L_vs_Drat} the relation between relative gas
content and the \HI-to-optical diameter ratio.
This figure shows that if a galaxy has a relatively high gas content,
it is because its gas disk is relatively extended; thus the average
\HI\ surface density will still be comparable to that in more gas-poor  
systems. 
It is, however, still unclear why some galaxies have much larger \HI\
disks than others (i.e.\ are much more gas-rich, cf.\
Sects.~\ref{subsec:HImass} and \ref{subsec:HIradii}).

\subsection{holes, rings and the relation with star formation}
\label{subsec:holes&rings}
Most early-type disk galaxies presented here whose \HI\ is
concentrated in a regular disk, have a central hole or depression in
their \HI\ distribution. 
Galaxies where we do not observe a central hole are either not well
resolved, or do not have a regularly rotating disk; it is likely
therefore, that {\it all} early-type galaxies with a regular gas disk
have a central depression in their \HI\ distribution.  
 
\begin{figure*}[htb] 
  \centerline{\psfig{figure=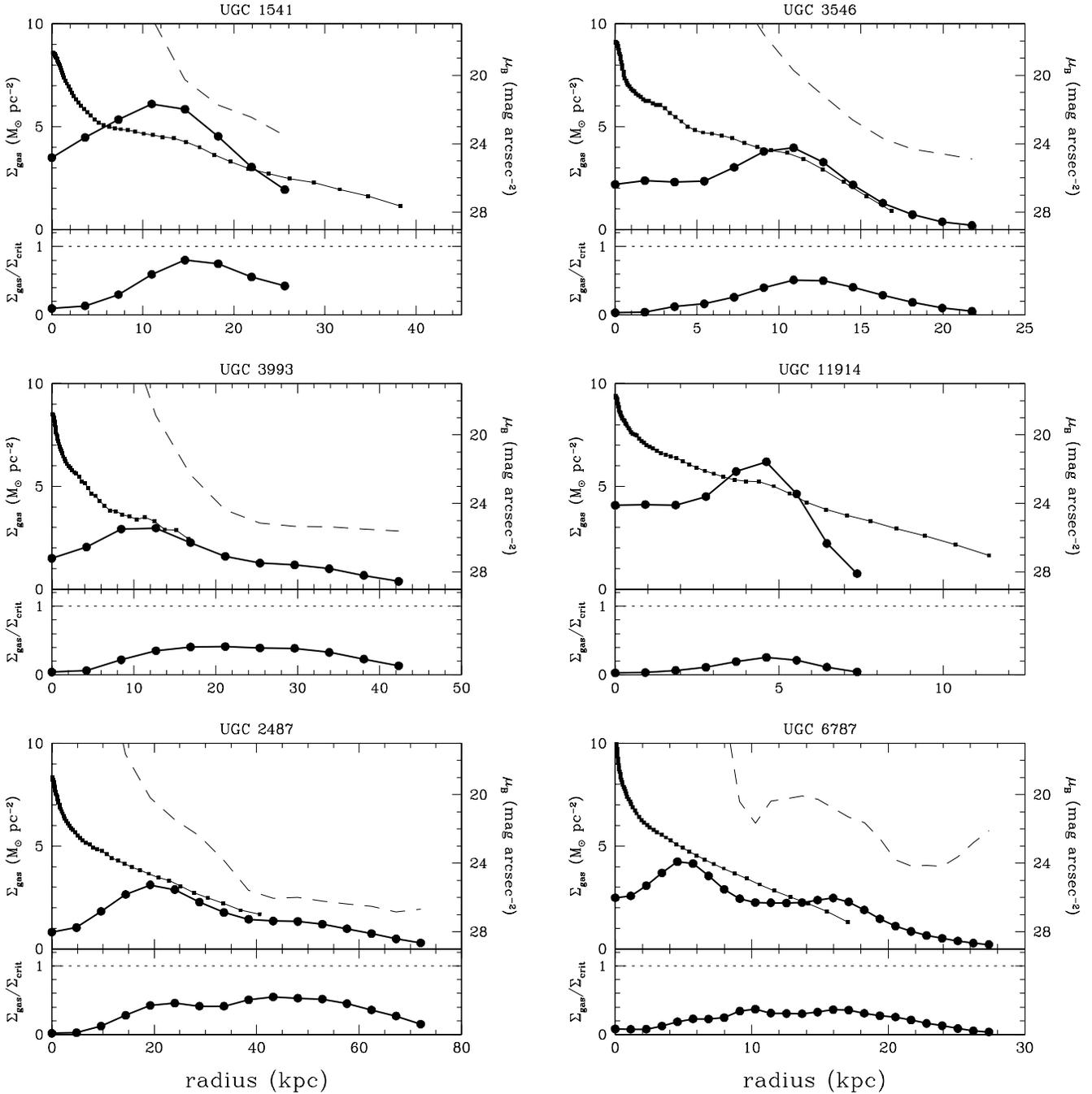,height=7in}}
  \caption{Comparison between gas and stellar light distributions in
  galaxies with pronounced gas rings. UGC~1541 and 3546 are barred
  galaxies; the others have no bars.
  {\it Top panels}: Filled circles and bold lines show the \HI\
  surface density profiles, multiplied by 1.44 to account for the presence of
  helium. Radii are converted to kpc. Filled squares and thin lines
  show B-band photometric profiles. Dashed lines show the critical
  density for star formation, according to
  \protect\citet{Kennicutt89}. For UGC~11914, the critical density is
  larger than 10~\msunpc2\ everywhere. 
  {\it Bottom panels}: Ratio between observed gas density and critical 
  density. The dotted line indicates the threshold for star formation.}   
  \label{fig:HIrings}
\end{figure*}
In some galaxies, most of the gas is concentrated in rings.
In most cases, the rings follow the orientation of the optical
disks, but there are also cases (e.g.\ UGC~12276) where the
orientation of the ring is different from that inferred from the
optical image or the velocity field.
Galaxies with gas rings offer an interesting possibility to study the   
relation between neutral gas and the stellar light distribution. 
In Fig.~\ref{fig:HIrings} we compare the radial gas distribution for  
6~galaxies with distinct gas rings with their B-band luminosity
profile. 
The gas surface density profiles are derived from the \HI\ profiles shown in
Appendix~\ref{app:atlas}, multiplied by 1.44 to account for the
presence of other elements (mostly helium). 
The photometric profiles are taken from a parallel study of the
stellar mass distribution in a subsample of the galaxies presented
here \citetext{Noordermeer et al., in prep.}. 

UGC~1541 and 3546 (top panels) are barred galaxies with distinct
spiral arms. 
In both cases there is a slight overdensity of blue light at the same
radius as the gas ring; this may simply reflect the fact that gas and
stars respond in the same way to the gravitational perturbations from
the bar and spiral arms. 

The other four galaxies, however, are not barred, and have very little
spiral structure. 
Yet in two cases, UGC~3993 and 11914 (middle panels), the \HI\ surface density
peaks correspond to small but distinct overdensities of light at the
same radii.  
In UGC~11914 this excess light is related to a prominent ring of blue
stars in the original images \citep[see also][]{Buta95}. 
H$\alpha$ emission is detected in this ring as well \citep{Pogge89b},
indicating that the gas is actively forming stars
\citep{Battinelli00}.  
We are not aware of H$\alpha$ observations of UGC~3993, but the stars
in its ring are significantly bluer than in the central parts,
indicating that a population of young stars may exist in this ring as 
well. 
In contrast, in UGC~2487 and 6787 (bottom panels), no excess light
over the regular exponential disks is detected and the colors of the
stars do not vary over the ring. 
\citet{Pogge93} found no H$\alpha$ emission in UGC~2487.  
 
In an attempt to understand why there is no current star formation in
the rings of UGC~2487 and 6787, while there are strong indications
that it {\it is}  happening in UGC~3993 and 11914 where the total gas
densities are similar, we compared the observed gas densities to the
threshold density for star formation derived by \citet{Kennicutt89}.  
Assuming that star formation in spiral galaxies is regulated by the
onset of gravitational instabilities in their gas disks,
\citeauthor{Kennicutt89} applied the criterion for disk stability from
\citet{Toomre64} to these systems and derived the following threshold
density for star formation:  
\begin{equation}
\Sigma_{\mathrm {crit}} = \alpha \frac{\kappa c}{3.36 G}.
\label{eq:sigmacrit}
\end{equation}
$\kappa$ is the epicyclic frequency, which we derived from rotation
curves from tilted ring fits to the velocity fields. 
For $c$, the velocity dispersion of the gas, we followed
\citeauthor{Kennicutt89}, and assumed  a constant value of 6 km~s$^{-1}$. 
For the high surface-brightness galaxies studied here, this value is
probably on the low side, and the actual values of $\Sigma_{\mathrm
  {crit}}$ may be higher than what we derive here. However, higher
values for the threshold density will only aggravate the discrepancy
described below, and we choose $c = 6$~km~s$^{-1}$ to be able to compare our
results directly with \citeauthor{Kennicutt89}.
$\alpha$ is a dimensionless quantity of order unity; from an empirical  
study of star formation cutoffs in spiral galaxies,
\citeauthor{Kennicutt89} derived a value of $\alpha = 0.67$. 
The applicability of this recipe was recently confirmed by
\citet{Martin01} for a sample of 32 nearby spiral galaxies. 
The critical densities in our galaxies are plotted, as a function of
radius, with the dashed lines in Fig.~\ref{fig:HIrings}.  
Due to the large rotation velocities and the correspondingly large
$\kappa$, the critical densities are high and the observed gas
densities are far below the threshold in all cases. 

Thus, if the description of star formation thresholds by
\citet{Kennicutt89} is correct, it is not surprising that we do not
see a correlation between the \HI\ rings and the stellar light
distribution in UGC~2487 and 6787; the gas densities are simply much 
too low to sustain large-scale star formation. 
But then it is unclear why UGC~3993 and 11914 do form stars.   
In fact, the galaxy out of these 6 which has the highest star
formation activity, UGC~11914, has the {\it lowest} ratio ($< 0.25$
everywhere) of observed to critical density!  

Note that we have ignored the contribution of molecular gas to the
total gas densities. 
It seems, however, unlikely that the presence of molecular gas can
explain the discrepancy between observed and critical gas densities in
these galaxies. 
Several CO-surveys have been carried out in recent years, and all find
that molecular gas is concentrated in the inner parts of galaxies
\citep{Sofue95,Wong02,Sofue03}, with exponential scale lengths
comparable to those of the underlying stellar disks \citep{Regan01}.    
No substantial column densities of CO emission are generally observed
at large radii. 
Thus, while we may have underestimated the total gas densities in the
inner regions of the galaxies in Fig.~\ref{fig:HIrings}, the effect
will be small in the \HI\ rings of interest, which all lie at large
radii, and the total gas densities are truly smaller than the critical
density. 

In recent years, more cases have been reported of star formation in
regions where the gas density is lower than the critical density from
equation~\ref{eq:sigmacrit}. 
\citet{Kennicutt89} and \citet{Martin01} already noted that a
fraction of the galaxies in their sample (notably M33 and NGC~2043)
showed widespread star formation, even though their gas surface
densities were below the threshold. 
\citet{vanZee97} observed 11 dwarf galaxies and showed that, although
the gas densities were below the threshold in all cases, some of them
had a substantial level of star formation.  
Several recent papers report the detection of star formation in the
outskirts of nearby galaxies, where the observed gas densities are
well below the threshold densities for star formation too
\citep{Ferguson98, Cuillandre01, Thilker05}. 
Thus, a number of cases have now been identified where the simple
criterion of \citet{Kennicutt89} does not apply. 

Recently, \citet{Schaye04} argued that, rather than being regulated by
global, gravitational instabilities, star formation requires a phase 
transition from warm to cold gas, and is thus governed by thermal
instabilities instead. 
He predicted that the critical density for star formation is more or
less independent of global properties such as rotation velocity, and
has a universal value in the range 3 -- 10 \msunpc2. 
Although it is not clear how this prediction compares to the large
spread in observed threshold densities from \citet{Kennicutt89}, it is
suggestive that all galaxies in Fig.~\ref{fig:HIrings} have gas
densities around the lower end of this range, and would thus be at the
verge of instability for star formation. 

It seems, however, premature to completely abandon
\citeauthor{Kennicutt89}'s theory, until more knowledge is obtained
about the small-scale distribution of the gas
\citep[cf.][]{vanZee97}. 
\citet{Braun97} studied the structure of \HI\ disks in 11
well-resolved, nearby galaxies and found that the \HI\ was distributed 
in small, high-density regions with covering factors between 6 and
50\%.  
Thus, if UGC~3993 and 11914 are extreme cases where the covering
factor is low, and UGC~2487 and 6787 have a smoother \HI\
distribution, the observed (lack of) correlation between \HI\
distribution and star formation in these galaxies might still be
consistent with the predictions from \citet{Kennicutt89}. 
Note, however, that \citeauthor{Kennicutt89} did not explicitly
account for the `patchiness' of gas in his galaxies; an average amount
of substructure in the gas distribution must implicitly be included in
his parameter $\alpha$.   
There is no a priori reason to assume that galaxies such as UGC~3993
or 11914 have much lower covering factors than average, but we cannot 
rule out this possibility, and high-resolution observations are needed 
to check this.  

\subsection{lopsidedness}
\label{subsec:lopsidedness}
Asymmetries in the morphology and kinematics of galaxies are still a
subject of debate. \citet{Rix95} found that about one third of all
spiral galaxies show large-scale asymmetries in their optical images.
Global \HI-profiles have asymmetric shapes for at least 50\% of
all disk galaxies \citep{Baldwin80, Richter94, Haynes98}, but without
a full 2D mapping of the \HI\ component in these galaxies, it is
difficult to determine the origin of the asymmetries in the profiles.  
With the advent of a large number of spatially resolved \HI\ images
and velocity fields, it was found that not only the spatial
distribution of the gas is often asymmetric, but that many galaxies
are also lopsided in their kinematics: \citet{Swaters99} estimated
that about half of all disk galaxies have asymmetric kinematics.  
Lopsidedness seems particularly common in late-type spirals;
\citet{Matthews98} found that out of a sample of 30 extremely
late-type galaxies, about 75\% had clearly lopsided \HI\ profiles.   

In some cases, lopsidedness can be linked to ongoing interactions with
companion galaxies.
However, lopsidedness is also common in isolated galaxies
\citep{Wilcots04}, implying that lopsidedness may be an intrinsic
feature of disk galaxies and that the underlying dark matter
distribution may be asymmetric too \citep{Jog97, Noordermeer01}.  
Even if lopsidedness in galaxies is triggered by tidal interactions or 
mergers, it must be a long-lived phenomenon to explain the relatively
high frequency of lopsidedness in isolated systems. 

A full, quantitative analysis of the occurrence of lopsidedness in the
galaxies in our sample is beyond the scope of the present paper. 
Instead, we have determined by eye how many of our early-type galaxies
are lopsided, and briefly discuss the results here. 
The degrees of asymmetry in the global profile, morphological
appearance and kinematical structure (velocity field and major axis
xv-diagram) are indicated with stars in Table~\ref{table:data}, where
0, 1, 2 or 3~stars mean not, mildly, moderately or severely lopsided
respectively.  
In some cases the gas is too strongly distorted or too poorly resolved
to determine whether it is asymmetric.  
These cases are indicated with n.a. (not available).   

Asymmetries in the global profiles and gas distribution are very
common in the early-type disk galaxies studied here. 
Respectively 35 and 34 galaxies (51 and 50\%) have at least mildly
asymmetric global profiles or surface density maps. 
In most cases, however, the asymmetries can be related  to the
presence of or interactions with companion galaxies. 
{\it All} galaxies, except one (UGC~5060), which have severely
lopsided gas distributions, and most cases where the gas distribution
is moderately lopsided, show clear signs of ongoing interactions. 
On the other hand, morphological lopsidedness is rare in early-type
disk galaxies without nearby companions or signs of recent
interactions. 
Of all 41 galaxies that do not show clear signs of interaction, only
15 (37\%) are morphologically asymmetric, usually only mildly. 
Only 3 isolated galaxies (7\%) are more than mildly asymmetric.   

Kinematical lopsidedness is rare in early-type galaxies in general,
interacting or not. 
Out of all the galaxies for which we can determine the degree of
symmetry in the kinematics (53), UGC~624 is the only one with a
severely lopsided velocity field; in this case the lopsidedness can 
be explained as a result of tidal interaction with its nearby
companion UGC~623.  
The only two cases of isolated galaxies with significantly lopsided
kinematics, UGC~5960 and UGC~11852, are both extreme galaxies in our
sample in the sense that the former is one of the least luminous
galaxies in our sample (${\mathrm {M_B}} = -17.39$), while the latter
has the most extended gas disk, relative to the optical diameter, of
all our sample galaxies (${\mathrm D}_\HI/{\mathrm D}_{25}^{B,c} =
4.1$).   

We conclude that lopsidedness in the \HI\ distribution is at least as
common in early-type disk galaxies as in later-type galaxies; truly
symmetric gas disks are rare.
In most cases, however, morphological asymmetries are clearly related
to ongoing interactions or accretion.
Morphological lopsidedness in isolated early-type disks, and
kinematical asymmetries, are much rarer, supporting the findings
of \citet{Matthews98}.

\section{notes on individual galaxies}
\label{sec:notes}
Many galaxies in our sample show interesting features in their
\HI\ distribution or kinematics which deserve special emphasis.
We discuss these below. \\[0.1cm] 
{\bf UGC~89} (NGC~23) and {\bf UGC~94} (NGC~26) are separated by only
10\arcmin. 
They form a loose group together with Scd galaxy UGC~79 (20\arcmin\ to
the southwest) and 4 other galaxies \citep{Garcia93}.  
A few small \HI\ clouds are visible in the neighbourhood. 
Optical counterparts are visible on the DSS for most of them, so they
are probably dwarf companions of the main group members. 
A tidal tail-like structure is visible south-east of {\bf UGC~94},
indicating a recent interaction or merger event. In our low-resolution
data cube, this tidal feature is seen to be connected to the main disk 
of {\bf UGC~94} and extends about 10\arcmin\ ($\sim 180$ kpc) to the
east. \\[0.1cm] 
{\bf UGC~499} (NGC~262), also known as Markarian~348, is a well-known
Seyfert~2 galaxy \citep{Koski78}. 
We detect \HI\ emission out to distances of 100~kpc from the center,
but it is clearly distorted and does not rotate regularly around the
center. 
{\bf UGC~508} (NGC~266) lies about 23\arcmin\ to the northeast of {\bf
UGC~499}; the peculiar structure of the gas in {\bf UGC~499} might be
a result of a past interaction with its neighbour. 
The \HI\ properties of {\bf UGC~499} were discussed in detail by
\citet{Heckman82} and \citet{Simkin87}. \\[0.1cm]  
{\bf UGC~624} (NGC~338) is one of the most lopsided galaxies in our
sample, both in the gas distribution and in the kinematics. 
The asymmetry could be a result of tidal interaction with nearby
neighbour UGC~623.  
Note also the small `blob' of gas just southeast of the main gas
disk. \\[0.1cm]
{\bf UGC~1310} (NGC~694) resides in a small group of galaxies, the main 
other members of which are NGC~680, 691 and 697 \citep{Garcia93}. 
The central regions of this group were imaged in \HI\ before by
\citet{VanMoorsel88}. Apart from UGC~1310, we also detect \HI\
emission in NGC~691, 697 and UGC~1313, but as these are all of later
morphological type, they are not included in the present study.
The \HI\ disk of {\bf UGC~1310} is barely resolved in our
observations. As a result, the line profiles are strongly affected by
beam-smearing; the method described in Sect.~\ref{subsec:vfields}
to derive the velocity field is therefore inadequate and fails to
recover the projected rotational velocities of the gas. \\[0.1cm]    
Most of the gas in {\bf UGC~2045} (NGC~972) seems to be concentrated
in a regularly rotating disk that has the same orientation as
dustlanes in the optical image. 
The line-profiles in this disk have complicated shapes and the method
described in Sect.~\ref{subsec:vfields} for the derivation of the
velocity field fails to recover the projected rotational velocities of
the gas. 
In the outer parts, several large filaments of gas are detected; the
gas in these filaments clearly does not follow the rotation of the
inner parts. \\[0.1cm]
{\bf UGC~2154} (NGC~1023) is the brightest galaxy of a nearby group of
13 galaxies \citep{Tully80}. The total \HI-flux for this galaxy is
quite large (80.13 Jy~km~s$^{-1}$), but the gas is scattered over
a large area, has a very low column-density and does not reside in a
regular disk. It seems most likely the result of a recent merger or
accretion event, possibly with one of the other group members
\citep[cf.][]{Sancisi84}. \\[0.1cm]  
Most of the gas in {\bf UGC~2183} (NGC~1056) is concentrated in an
edge-on disk which coincides with the dust lane seen in the optical
image.  
Due to this edge-on orientation, projection effects lead to strongly
non-Gaussian line profiles; the method described in
Sect.~\ref{subsec:vfields} to derive the velocity field is therefore 
inadequate and fails to recover the projected rotational velocities of
the gas in the central parts. 
At larger radii, the gas seems to warp out of the plane and the 
orientation becomes more face-on. \\[0.1cm]
{\bf UGC~2487} (NGC~1167) is a nice example of an S0 galaxy with an
extended and regularly rotating \HI\ disk. 
The gas seems to be in circular motion at radii up till 80 kpc.   
The central hole does not reflect a true absence of gas, but is rather
the result of absorption against a central continuum source. \\[0.1cm]   
The companion east of {\bf UGC~2916} is PGC~14370. When the data are 
smoothed to lower resolution, some emission is seen to bridge the
space between the two galaxies, indicating that some interaction is
going on. There are clear signs of off-planar gas in PGC~14370, but
the main disk of {\bf UGC~2916} seems relatively
undisturbed. \\[0.1cm]  
{\bf UGC~2953} (IC~356) is by far the best resolved galaxy in our
sample.  
The gas is concentrated in pronounced spiral arms that extend far
beyond the bright optical disk. 
The velocity field shows distinctive `wiggles' in the isovelocity
contours where the gas crosses the arms. \\[0.1cm] 
The \HI\ distribution, velocity field and global profile of {\bf
UGC~3205} are almost perfectly symmetric, except for the twisting of
the isovelocity contours in the bar region. \\[0.1cm]
{\bf UGC~3354} is almost perfectly edge-on and has a peculiar double
warp. Both the stellar and gas disks seem to warp `clockwise'
first. At larger radii, where no starlight can be detected, the gas
disk reverses its warp to the opposite direction.
Projection effects lead to complicated line-profiles in the inner
parts of the disk; the method described in Sect.~\ref{subsec:vfields}
to derive the velocity field is inadequate there and fails to recover
the projected rotational velocities of the gas.  \\[0.1cm]
The gas distribution in {\bf UGC~3426} is irregular and seems to be
connected to the gas disk of UGC~3422, $\approx 100$ kpc to the
north-west. The projected surface density of the gas is very low, with
a maximum of only about 0.3~\msunpc2. It appears as if the gas in {\bf
  UGC~3426} has been tidally drawn out of the gas disk of
UGC~3422. \\[0.1cm] 
{\bf UGC~3642} is a very peculiar galaxy. The gas inside the bright
optical disk seems to be regularly rotating in the same plane as
defined by the stellar light distribution.  But further out, the
direction of motion is reversed. Each component individually appears
undisturbed and the gas seems to be in regular rotation in both the
inner and the outer regions. Two options seem possible to explain the
peculiar kinematics of this galaxy. The first is that both components
are fully decoupled, in which case this galaxy would carry some
resemblance with the `Evil Eye' galaxy (NGC~4826), which also has two
decoupled gas disks \citep{Braun92,Rubin94}. However, both the total
\HI\ mass and the size of the disks in {\bf UGC~3642} are an order of
magnitude larger than in NGC~4826. Alternatively, the gas disk could
be extremely warped, such that the direction of motion along the line
of sight is reversed. We are unable to distinguish between the two
options, and more detailed observations and modelling are needed to
clarify the nature of this galaxy. Note that the total extent of the
outer gas disk is extremely large, with $D_{\HI} > 100 \, {\mathrm
  {kpc}}$. \\[0.1cm]     
At first sight, {\bf UGC~3965} (IC~2204) seems a similar case as
UGC~3642, with the kinematical position angle changing by almost
180\deg. But in this case, the observed velocity field can also be
explained by an almost face-on disk with a mild warp. While the gas
would be moving in the same direction everywhere, its radial velocity
along the line of sight could easily be reversed. \\[0.1cm] 
{\bf UGC~4605} (NGC~2654) is almost perfectly edge-on in the inner
parts, but the gas seems mildly warped in the outer parts. 
The position-velocity diagram shows that the rotation velocities
decline strongly towards the edge of the \HI\ disk; the decline starts
already before the onset of the warp and must reflect a true decrease
in the rotation velocities. 
In the very inner parts, the line-profiles are strongly affected by
projection effects and the method described in
Sect.~\ref{subsec:vfields} to derive the velocity field fails to 
recover the projected rotational velocities of the gas there.
\\[0.1cm] 
{\bf UGC~4637} (NGC~2655) is a nearby Seyfert 2 galaxy whose
nuclear regions show complex structure, both in high-resolution radio
continuum observations as in optical emission lines
\citep{Keel88}. These authors attribute the complexity of the central
parts to a recent interaction or merger event. This interpretation
seems to be confirmed by the large-scale structure of the neutral gas
seen in our observations. The \HI\ is clearly disturbed, with a large
extension to the northwest; the gas does seem to have a general sense
of rotation, but it is clearly not on regular circular orbits. The
optical image of this galaxy shows some distinct loops and shells,
further confirming the hypothesis of a recent interaction. \\[0.1cm] 
{\bf UGC~4666} (NGC~2685) is also known as the Spindle or Helix
galaxy, because of its prominent polar ring. Early \HI\ observations
of this galaxy were presented by \citet{Shane80}. Our observations, at
higher sensitivity and resolution, confirm both the outer gas ring as
well as \HI\ related to the helix-like structure inside this ring. The
gas in the helix is kinematically distinct from the rest of the galaxy
and seen at almost right angles to the outer ring. \\[0.1cm]       
{\bf UGC~4862} (NGC~2782) has a large, banana-shaped tail of gas
northwest of the main disk. The gas in the central regions is clearly
disturbed as well and does not show signs of regular rotation around
the center. The optical image is peculiar too, with some striking
shells. These facts were already noted by \citet{Smith94}, who
interpreted the peculiar structure as evidence for a recent merger of
the main galaxy with a low-mass companion. 
\citet{Schiminovich94,Schiminovich95} studied \HI\ in shells around 
elliptical galaxies. It seems not unreasonable to assume that {\bf
  UGC~4862} is a similar case as their galaxies, but in an earlier
stage of its evolution. \\[0.1cm]  
The main disk of {\bf UGC~5253} (NGC~2985) is regular in its
distribution and kinematics. 
In the outer parts, a large one-armed spiral of gas is seen to extend
towards a small `blob' of gas southwest of the galaxy. 
{\bf UGC~5253} lies about 20\arcmin\ west of the late-type spiral  
UGC~5316, which also has a disturbed morphology
\citep[e.g.][]{Vorontsov77}. The redshifts of the two systems differ 
by about 250~km~s$^{-1}$ only, so the peculiar structures of both
galaxies could well be the result of a tidal interaction. \\[0.1cm] 
{\bf UGC~5559} (NGC~3190) is part of Hickson compact group of galaxies
44, together with UGC~5554, 5556 and 5562 \citep{Hickson89}. All group
members lie in our field of view, but the observations for these
galaxies were done with the `old' Westerbork receivers and the
sensitivity is insufficient to detect any emission in UGC~5554 and
5562. {\bf UGC~5559} itself is only barely detected, and the
signal-to-noise ratio of the data is too low to make a detailed study
of the gas distribution in this galaxy.  
Only for the more gas-rich, late-type spiral UGC~5556 useful \HI\
surface density maps and velocity fields could be obtained, but as
this galaxy does not meet our selection criteria, it is not discussed
here further. \\[0.1cm] 
{\bf UGC~5906} (NGC~3380) is poorly resolved and the line-profiles
suffer from beam-smearing. 
As a result, the method described in Sect.~\ref{subsec:vfields}
to derive the velocity field is inadequate and fails to recover the
projected rotational velocities of the gas. \\[0.1cm]  
{\bf UGC~5960} (NGC~3413) is close to edge-on and projection effects
lead to non-Gaussian line profiles. As a result, the method described
in Sect.~\ref{subsec:vfields} to derive the velocity field is
inadequate and fails to recover the projected rotational velocities of
the gas. \\[0.1cm]  
Due to the poor resolution and resulting beam-smearing in our
observations of {\bf UGC~6001} (NGC~3442), the method described in   
Sect.~\ref{subsec:vfields} to derive the velocity field is inadequate
and fails to recover the projected rotational velocities of the gas.
\\[0.1cm]      
The central hole in the disk of {\bf UGC~6118} (NGC~3504) is not due
to a true absence of gas, but rather an artefact caused by \HI\
absorption against a central continuum source (see also
UGC~2487). \\[0.1cm]  
{\bf UGC~6621} (NGC~3786) and {\bf UGC~6623} (NGC~3788) are clearly
interacting. A giant tidal tail of \HI\ extends almost 50 kpc
northward from {\bf UGC~6623}. On our optical image, a
low-surface-brightness counterpart is visible.   
Although the gas of both galaxies seems to overlap in the space
between them, it is separated in velocity and we can distinguish which
gas belongs to which galaxy in the full resolution data cube. 
Thus we were able to generate radial profiles for the gas in each
galaxy separately.  
At 60\arcsec\ resolution, the data become too heavily smoothed and the
regions of emission from the individual galaxies start overlapping in
individual channel maps as well. 
We cannot make global profiles of the emission of each galaxy
separately anymore at this resolution, and we were forced to generate
the global profiles from the $30\arcsec$ data. 
But at $30\arcsec$ already the galaxies are hardly resolved and the
masks for this resolution do not miss much flux compared to those of
the $60\arcsec$ data. \\[0.1cm]    
{\bf UGC~6742} (NGC~3870) is poorly resolved and the line-profiles
suffer from beam-smearing. 
As a result, the method described in Sect.~\ref{subsec:vfields}
to derive the velocity field is inadequate and fails to recover the
projected rotational velocities of the gas. \\[0.1cm]  
The gas in the inner parts of {\bf UGC~6786} (NGC~3900) follows the
light distribution of the optical image. Most of the gas in the outer
parts is concentrated in two diffuse spiral arms which seem to be
warped with respect to the inner parts. \\[0.1cm] 
The well-studied galaxy {\bf UGC~7166} (NGC~4151) is one of the
original Seyfert galaxies \citep{Seyfert43}. The \HI\ in this 
galaxy follows the stellar light in the bar and the spiral arms
\citep[cf.][]{Pedlar92}. \\[0.1cm]
{\bf UGC~7256} (NGC~4203) has a peculiar, filamentary gas
distribution.  
The gas has a general sense of motion around the center, but it is
clearly not on regular circular orbits.  
This galaxy bears some resemblance with UGC~4637 (see
above). \\[0.1cm]  
{\bf UGC~7489} (NGC~4369) has the smallest \HI\ disk, relative to its
optical diameter, of all galaxies in our sample (${\mathrm
  D}_\HI/{\mathrm D}_{25}^{B,c} = 0.86$), with the exception of a few
interacting systems or galaxies where the \HI\ or optical radii are
not well defined (e.g.\ UGC~5559 or 10448). 
In the optical image, very weak spiral structure and dust absorption
can be seen at the locations of the \HI\ emission. \\[0.1cm] 
{\bf UGC~7704} (NGC~4509) has a peculiar morphology. The optical image
has a position angle of about 155\deg, but the \HI\ map seems to be
almost perpendicular to this, with a position angle of about 57\deg. 
The kinematical major axis is different again, about 37\deg. 
{\bf UGC~7704} does not seem to have any major companions which could
cause this strange structure, so the true nature of it remains
unclear. 
Due to the poor resolution, the line-profiles in this galaxy suffer
from beam-smearing and the method described in
Sect.~\ref{subsec:vfields} to derive the velocity field fails to 
recover the projected rotational velocities of the gas. \\[0.1cm] 
No \HI\ gas is detected in the bulge of {\bf UGC~7989} (NGC~4725), nor
in the giant bar. 
It is rather concentrated in narrow spiral arms, which coincide with
the stellar arms. 
The outer arm on the east side is particularly prominent, whereas the
arms on the northwest side seem to be truncated, causing marked
asymmetries in the \HI\ surface density map and global
profile. \\[0.1cm]   
Most of the gas in {\bf UGC~8271} (NGC~5014) seems to be concentrated
in a ring which is tilted with respect to the main stellar disk. 
Close inspection of the optical image reveals a faint polar ring-like 
structure, and the \HI\ at the corresponding locations seems to be 
rotating in the same plane as defined by this ring. A tail of gas
extends further to the south. 12\arcmin\ to the north-northeast, a 
number of large \HI\ clouds are detected, all without optical
counterparts. {\bf UGC~8271} resides in a loose group, the main other 
members of which are NGC~5005 and 5033. The peculiar phenomena in the
gas distribution and kinematics in and around {\bf UGC~8271} may all
be related and suggest a recent merger or accretion event, possibly
with a small gas-rich member of the same group of galaxies.  
Due to the complicated structure of the gas in this galaxy and the
resulting projection effects, the line-profiles are strongly
non-Gaussian and the method described in Sect.~\ref{subsec:vfields} to
derive the velocity field fails to recover the projected  velocities
of the gas. \\[0.1cm]
{\bf UGC~8805} (NGC~5347) is at first sight a similar case as
UGC~7704. The position angle of the \HI-disk seems to be almost 
perpendicular to that of the optical disk. Note though that we only
get sufficient signal at 60\arcsec\ resolution; the optical disk then
fits almost entirely in one resolution element. There is a hint that
the \HI\ is actually elongated along the optical image in the central
regions, but observations at higher resolution and sensitivity are
required to confirm this. If indeed the gas is aligned with the stars
in the inner parts, {\bf UGC~8805} could be an example of a galaxy
with an extreme warp. This would also explain the strong decline of
radial velocities along the major axis, as shown in the
xv-slice. \\[0.1cm] 
The optical image of {\bf UGC~8863} (NGC~5377) is completely dominated
by a giant bar. 
Two faint spiral arms are seen to extend from its edges. The \HI\
emission nicely follows the bar and spiral arms; the latter are much
more pronounced here.  
In the center, bar-induced streaming motions produce a twist in the
isovelocity contours.\\[0.1cm] 
{\bf UGC~9133} (NGC~5533) has a large one-armed spiral in the outer
regions which seems to be warped with respect to the main disk. 
The inner parts are highly regular. 
The xv-diagram indicates that the rotation velocities are declining. 
In the outer parts, this can be explained as a result of the warp, but
the decline sets in well within the radius of the warp and must
reflect a truly falling rotation curve. \\[0.1cm]
The \HI\ which we detect in {\bf UGC~10448} (NGC~6186) is very
peculiar.   
The gas seems offset from the optical image by about 30\arcsec.  
Close inspection of the optical image reveals a faint galaxy behind 
the main disk.  
Thus, we are seeing here the coincidental alignment of two spiral
galaxies.  
This is further confirmed by the redshift of the \HI\ emission of
about 11350 km~s$^{-1}$, whereas the optical redshift of the
foreground galaxy is 2935 km~s$^{-1}$ \citep[UZC;][]{UZC}.
The foreground galaxy has been detected at the correct redshift in
\HI\ as well \citep{Rosenberg00}, but the UGC contains the redshift of
the background system and confused us into observing at the wrong 
frequency.  
As no morphological classification, nor photometric data, are 
available for the background galaxy, we have excluded this galaxy from 
the statistical analysis in Sect.~\ref{sec:results}, and merely show
the data for general interest. \\[0.1cm]    
{\bf UGC~11269} (NGC~6667) must recently have undergone a merger. 
A giant one-armed spiral arm seems to extend out from the main
disk. It is partly visible in the optical image as well. 
Most of this material seems to be virialized already, since it follows
closely the velocities of the gas in the inner parts. 
Note also the peculiar, asymmetric shape of the position-velocity
diagram along the major axis: at the receding side, the peak velocity
is much higher than at the approaching side. 
On both sides, the rotation velocities decline strongly away from the
center. \\[0.1cm]  
Most of the gas in {\bf UGC~11670} (NGC~7013) is concentrated in the
bar and the spiral arms. 
In the outer regions, the gas distribution is irregular, with several
filaments emanating from the main disk.
The line profiles in the inner parts are strongly non-Gaussian; the
method described in Sect.~\ref{subsec:vfields} to derive the velocity
field is inadequate there and fails to recover the projected
rotational velocities of the gas. \\[0.1cm]   
The gas in {\bf UGC~11914} (NGC~7217) is predominantly concentrated in
a ring. 
The stellar population in the ring is distinctively bluer than the
surroundings, indicating that the gas is associated with an enhanced
level of star formation (see Sect.~\ref{subsec:holes&rings}). 
\\[0.1cm]     
Most of the gas in {\bf UGC~12276} (NGC~7440) is concentrated in a
ring which is elongated perpendicular to the optical and kinematical
major axis. It seems unlikely that this peculiar geometry can be
explained solely by the presence of the little companion
$\approx$~40~kpc to the east, so the true nature of it remains
unclear. \\[0.1cm] 
Due to the poor resolution and resulting beam-smearing in the
observations of {\bf UGC~12713}, the method described in
Sect.~\ref{subsec:vfields} to derive the velocity field is inadequate
and fails to recover the projected rotational velocities of the
gas. \\[0.1cm] 
{\bf UGC~12815} (NGC~7771) is strongly distorted and seems to be
interacting with its neighbours UGC~12813, NGC~7771A and UGC~12808. 
Two giant tidal tails extend to more than 100~kpc away from the
system.

\section{Conclusions}
\label{sec:conclusions}
In the previous sections, we have presented the results of a study
of the \HI\ properties of a sample of 68 early-type disk galaxies,
with morphological type ranging from S0 to Sab and absolute B-band
magnitude between -14 and -22. A number of conclusions can be drawn: 
\begin{itemize}
\item The \HI\ content of early-type disk galaxies is highly
  variable. There is a wide range in \HI\ mass-to-light ratios
  ${\mathrm M}_\HI / {\mathrm L}_B$, with the most \HI\ rich galaxies 
  in our sample containing about 2 orders of magnitude more gas,
  relative to the stellar luminosity, than the most gas-poor systems.    
  The average values for $\log({\mathrm M}_\HI / {\mathrm L}_B)$
  are $-0.62 \pm 0.44$ and $-0.50 \pm 0.40$ in solar units for Sa/Sab
  and S0/S0a galaxies respectively.  
  The errors give the standard deviations of the distributions.
\item The average ratio ${\mathrm D}_\HI / {\mathrm D}_{25}^{B,c}$
  between \HI\ and optical diameter, defined at 1~\msunpc2\ and 25
  B-band\magasas\ respectively, is $1.72 \pm 0.70$ for Sa/Sab
  galaxies and $2.11 \pm 0.70$ for S0/S0a's. These values are
  comparable to those observed for later-type spiral galaxies, but
  with larger spread.  
\item The average \HI\ surface brightness in our sample galaxies is
  slightly lower than that in later-type galaxies. 
  Within our sample, the variations from galaxy to galaxy are small. 
  The average values for \mbox{$<\!\Sigma_\HI\!>_{{\mathrm R}_\HI}$},
  the effective \HI\ surface brightness inside the 1~\msunpc2\
  isophote, are $2.8 \pm 0.8$ and $2.9 \pm 1.4$ for Sa/Sab and S0/S0a
  galaxies respectively. 
\item All early-type galaxies whose gas is distributed in a regular
  rotating disk have a central hole or depression in their \HI\
  distribution. 
\item A number of galaxies in our sample have distinct, axisymmetric
  rings of gas. In some of these cases, the overdensity of gas
  coincides with regions of enhanced star formation and a population
  of young stars, even though the gas densities are far below the
  threshold for star formation derived by \citet{Kennicutt89}. 
  In other cases no star formation activity is present, even though
  the gas densities are comparable. 
  These discrepancies suggest the existence of an additional
  regulation mechanism for star formation at low gas densities, the
  exact nature of which still needs to be clarified. 
\item Morphological and kinematical peculiarities are very common in
  early-type disk galaxies, often related to ongoing or recent
  interaction events. In many galaxies, we see indications for a tidal
  origin of the gas. Interactions with neighbour galaxies seem to be a
  driving force in the evolution of the neutral gas component in many
  early-type disk galaxies.   
\item Many early-type disk galaxies have lopsided morphologies. In
  most cases the asymmetries can be explained as the result of
  interaction, accretion or merger events. Few isolated galaxies have
  lopsided gas distributions. 
\item Kinematic lopsidedness is rare in early-type disk galaxies, even 
  in interacting systems.
\end{itemize}

The data presented in the atlas in the appendix form the basis for a
study of the rotation curves and dark matter content in these
early-type disk galaxies. This study is currently ongoing, and the
results will be presented in a forthcoming paper.

\begin{acknowledgements}
Great appreciation goes to the members of the WHISP team, in
particular Yuan Tang, for carrying out the crucial first stages of the 
data reduction of the WHISP observations.  
EN is grateful to the Department of Physics and Astronomy of the
University of Sheffield for the hospitality during part of this work.    
The Westerbork Synthesis Radio Telescope is operated by ASTRON
(Netherlands Foundation for Research in Astronomy) with support from
the Netherlands Foundation for Scientific Research (NWO).  
We made extensive use of data from the Lyon Extragalactic Database
(LEDA; \mbox{http://leda.univ-lyon1.fr/}) and the NASA/IPAC
Extragalactic Database (NED; http://nedwww.ipac.caltech.edu/) which is
operated by the Jet Propulsion Laboratory, California Institute of
Technology, under contract with the National Aeronautics and Space
Administration. 
This research was funded by NWO grant no.\ 614031009. 
\end{acknowledgements}

\bibliographystyle{aa}
\bibliography{../../../thesis/abbrev,../../../thesis/refs}

\clearpage
\onecolumn


\LTcapwidth=7.25in

\small{
\begin{longtable}{rl@{\hspace{0.6cm}}r@{\hspace{0.25cm}}r@{\hspace{0.25cm}}r@{\hspace{0.6cm}}r@{\hspace{0.25cm}}r@{\hspace{0.1cm}}r@{\hspace{0.7cm}}lrrrll}

  \caption{WHISP early-type disk galaxy sample: basic data. (1)~UGC
  number, (2)~alternative name, (3)~RA, (4)~Dec, (5)~morphological
  type, (6)~absolute B-band magnitude, (7)~heliocentric radial
  velocity, (8)~distance, (9)~inclination angle (superscripts
  indicate source: $^t$~tilted ring fits to velocity fields, $^L$~LEDA  
  and $^e$~estimated from our optical data) and (10)~group/cluster
  membership (superscripts indicate source: $^1$~\citet{Garcia93},
  $^2$~\citet{Tully88}, $^3$~\citet{Tully80}, $^4$~\citet{Hickson89}
  and $^5$ this study).  
  Columns (3)~--~(5) were taken from NED, column (6) from LEDA and 
  columns (7) and (8) from this study (cf.\ Table~\ref{table:data}).  
\label{table:basicdata}} \\
    
  \hline \hline 
  \multicolumn{1}{c}{UGC} &
  \multicolumn{1}{c@{\hspace{0.6cm}}}{alternative} &  
  \multicolumn{3}{c@{\hspace{0.6cm}}}{RA (2000)} &
  \multicolumn{3}{c@{\hspace{0.7cm}}}{Dec (2000)} &  
  \multicolumn{1}{c@{\hspace{0.7cm}}}{Type} &
  \multicolumn{1}{c}{M$_{\mathrm B}$} &
  \multicolumn{1}{c}{V$_{\mathrm{hel}}$} & \multicolumn{1}{c}{$D$} & 
  \multicolumn{1}{c}{$i$} & \multicolumn{1}{c}{\hspace{-1.cm}group/cluster} \\ 
  
  & \multicolumn{1}{c@{\hspace{0.6cm}}}{name} & 
  \multicolumn{1}{c@{\hspace{0.3cm}}}{\it h} &
  \multicolumn{1}{c@{\hspace{0.25cm}}}{\it m} & 
  \multicolumn{1}{c@{\hspace{0.7cm}}}{\it s} & 
  \multicolumn{1}{c@{\hspace{0.35cm}}}{$^{\circ}$} &
  \multicolumn{1}{c@{\hspace{0.25cm}}}{\arcmin} & 
  \multicolumn{1}{c@{\hspace{0.7cm}}}{\arcsec} & &
  \multicolumn{1}{c}{mag} & \multicolumn{1}{c}{km~s$^{-1}$} &
  \multicolumn{1}{c}{Mpc} & \multicolumn{1}{c}{\deg} &
  \multicolumn{1}{c}{\hspace{-1.cm}membership} \\       
    
  \multicolumn{1}{c}{(1)} & \multicolumn{1}{c@{\hspace{0.6cm}}}{(2)} &  
  \multicolumn{3}{c@{\hspace{0.6cm}}}{(3)} &
  \multicolumn{3}{c@{\hspace{0.7cm}}}{(4)} &  
  \multicolumn{1}{c@{\hspace{0.7cm}}}{(5)} & \multicolumn{1}{c}{(6)} & 
  \multicolumn{1}{c}{(7)} & \multicolumn{1}{c}{(8)} &
  \multicolumn{1}{c}{(9)} & \multicolumn{1}{c}{\hspace{-1.cm}(10)}\\ 
  \hline 
  \endfirsthead

  \caption[]{basic data: continued}\\
  \hline \hline
  \multicolumn{1}{c}{UGC} &
  \multicolumn{1}{c@{\hspace{0.6cm}}}{alternative} &  
  \multicolumn{3}{c@{\hspace{0.6cm}}}{RA (2000)} &
  \multicolumn{3}{c@{\hspace{0.7cm}}}{Dec (2000)} &  
  \multicolumn{1}{c@{\hspace{0.7cm}}}{Type} &
  \multicolumn{1}{c}{M$_{\mathrm B}$} &
  \multicolumn{1}{c}{V$_{\mathrm{hel}}$} & \multicolumn{1}{c}{$D$} &
  \multicolumn{1}{c}{$i$} & \multicolumn{1}{c}{\hspace{-1.cm}group/cluster} \\ 
    
  & \multicolumn{1}{c@{\hspace{0.6cm}}}{name} & 
  \multicolumn{1}{c@{\hspace{0.3cm}}}{\it h} &
  \multicolumn{1}{c@{\hspace{0.25cm}}}{\it m} & 
  \multicolumn{1}{c@{\hspace{0.7cm}}}{\it s} & 
  \multicolumn{1}{c@{\hspace{0.35cm}}}{$^{\circ}$} &
  \multicolumn{1}{c@{\hspace{0.25cm}}}{\arcmin} & 
  \multicolumn{1}{c@{\hspace{0.7cm}}}{\arcsec} & &
  \multicolumn{1}{c}{mag} & \multicolumn{1}{c}{km~s$^{-1}$} &
  \multicolumn{1}{c}{Mpc} & \multicolumn{1}{c}{\deg} &
  \multicolumn{1}{c}{\hspace{-1.cm}membership} \\      
    
  \multicolumn{1}{c}{(1)} & \multicolumn{1}{c@{\hspace{0.6cm}}}{(2)} &  
  \multicolumn{3}{c@{\hspace{0.6cm}}}{(3)} &
  \multicolumn{3}{c@{\hspace{0.7cm}}}{(4)} &  
  \multicolumn{1}{c@{\hspace{0.7cm}}}{(5)} & \multicolumn{1}{c}{(6)} &
  \multicolumn{1}{c}{(7)} & \multicolumn{1}{c}{(8)} &
  \multicolumn{1}{c}{(9)} & \multicolumn{1}{c}{\hspace{-1.cm}(10)}\\ 
  \hline 
  \endhead

  \hline
  \endfoot
  
  89     & NGC 23   & 0  &  9 & 53.4 & 25 & 55 & 26 & SB(s)a          & -21.48 &  4555  &  62.1 & 50$^t$ & LGG 2$^1$ \\ 
  94     & NGC 26   & 0  & 10 & 25.9 & 25 & 49 & 55 & SA(rs)ab 	      & -20.21 &  4589  &  62.6 & 42$^t$ & LGG 2$^1$ \\  
  232    & --       & 0  & 24 & 38.7 & 33 & 15 & 22 & SB(r)a 	      & -19.96 &  4839  &  66.3 & 48$^L$ & -- \\  
  499    & NGC 262  & 0  & 48 & 47.1 & 31 & 57 & 25 & SA(s)0/a:       & -20.09 &  4534  &  62.0 & 46$^L$ & LGG 14$^1$ \\  
  508    & NGC 266  & 0  & 49 & 47.8 & 32 & 16 & 40 & SB(rs)ab 	      & -21.97 &  4647  &  63.6 & 25$^t$ & LGG 14$^1$ \\  
  624    & NGC 338  & 1  &  0 & 36.4 & 30 & 40 & 8  & Sab 	      & -21.42 &  4772  &  65.1 & 59$^t$ & LGG 14$^1$ \\  
  798    & IC 1654  & 1  & 15 & 11.9 & 30 & 11 & 41 & (R)SB(r)a       & -19.67 &  4896  &  66.7 & 40$^t$ & LGG 18$^1$ \\  
  1310   & NGC 694  & 1  & 50 & 58.5 & 21 & 59 & 51 & S0? pec 	      & -19.23 &  2958  &  40.2 & 47$^e$ & LGG 34$^1$ \\  
  1541   & NGC 797  & 2  & 3  & 27.9 & 38 & 7  & 1  & SAB(s)a 	      & -21.10 &  5649  &  77.0 & 41$^t$ & pair with NGC~801$^5$ \\  
  2045   & NGC 972  & 2  & 34 & 13.4 & 29 & 18 & 41 & Sab             & -20.39 &  1527  &  21.4 & 61$^e$ & 52 -0$^2$ (Cetus) \\ 
  2141   & NGC 1012 & 2  & 39 & 14.9 & 30 & 9  & 6  & S0/a?           & -18.02 &  987   &  14.3 & 90$^e$ & 17 -3$^2$ (Triang.\ Spur) \\  
  2154   & NGC 1023 & 2  & 40 & 24.0 & 39 & 3  & 48 & SB(rs)0-        & -20.23 &  695   &  10.9 & 64$^L$ & N 1023 group$^3$ \\  
  2183   & NGC 1056 & 2  & 42 & 48.3 & 28 & 34 & 27 & Sa: 	      & -19.60 &  1540  &  21.5 & 90$^e$ & --  \\  
  2487   & NGC 1167 & 3  & 1  & 42.4 & 35 & 12 & 21 & SA0- 	      & -21.73 &  4950  &  67.4 & 36$^t$ & LGG 80$^1$ \\  
  2916   & --       & 4  & 2  & 33.8 & 71 & 42 & 21 & Sab 	      & -20.34 &  4518  &  63.5 & 46$^t$ & LGG 115$^1$ \\  
  2941   & IC 357   & 4  &  3 & 44.0 & 22 & 9  & 33 & SB(s)ab 	      & -21.38 &  6261  &  83.9 & 45$^e$ & group with UGC~2942/43$^5$ \\  
  2953   & IC 356   & 4  & 7  & 46.9 & 69 & 48 & 45 & SA(s)ab pec     & -21.49 &  894   &  15.1 & 50$^t$ & 12 +15$^2$ (U.\ Major) \\  
  3205   & --       & 4  & 56 & 14.8 & 30 &  3 & 8  & Sab	      & -21.47 &  3587  &  48.7 & 67$^t$ & -- \\  
  3354   & --       & 5  & 47 & 18.2 & 56 &  6 & 44 & Sab: 	      & -20.00 &  3084  &  43.6 & 90$^e$ & -- \\  
  3382   & --       & 5  & 59 & 47.7 & 62 &  9 & 29 & SB(rs)a 	      & -20.62 &  4501  &  62.8 & 16$^t$ & -- \\  
  3407   & --       & 6  & 9  & 8.1  & 42 &  5 & 7  & Sa 	      & -20.05 &  3606  &  49.8 & 45$^e$ & -- \\  
  3426   & --       & 6  & 15 & 36.3 & 71 & 2  & 15 & S0: 	      & -20.89 &  4005  &  56.6 & 37$^e$ & LGG 135$^1$ \\  
  3546   & NGC 2273 & 6  & 50 & 8.7  & 60 & 50 & 45 & SB(r)a 	      & -20.16 &  1838  &  27.3 & 52$^t$ & 24 -1$^2$ (Lynx) \\ 
  3580   & --       & 6  & 55 & 30.8 & 69 & 33 & 47 & SA(s)a pec:     & -18.19 &  1200  &  19.2 & 63$^t$ & 12 -0$^2$ (U.\ Major) \\  
  3642   & --       & 7  & 4  & 20.3 & 64 & 1  & 13 & SA0 	      & -21.00 &  4498  &  62.9 & 45$^t$ & LGG 140$^1$ \\  
  3965   & IC 2204  & 7  & 41 & 18.1 & 34 & 13 & 56 & (R)SB(r)ab      & -18.64 &  4588  &  62.5 & 10$^e$ & -- \\  
  3993   & --       & 7  & 55 & 44.0 & 84 & 55 & 35 & S0? 	      & -19.61 &  4366  &  61.9 & 20$^t$ & pair with UGC~3992$^5$ \\  
  4458   & NGC 2599 & 8  & 32 & 11.3 & 22 & 33 & 38 & SAa 	      & -21.12 &  4757  &  64.2 & 27$^t$ & pair with PGC~23972$^5$ \\  
  4605   & NGC 2654 & 8  & 49 & 11.9 & 60 & 13 & 16 & SBab: sp 	      & -20.08 &  1350  &  20.9 & 90$^e$ & 13 -6$^2$ (U. Major S.\ Spur) \\  
  4637   & NGC 2655 & 8  & 55 & 37.7 & 78 & 13 & 23 & SAB(s)0/a       & -20.97 &  1415  &  22.4 & 30$^e$ & 12 -10$^2$ (U. Major) \\  
  4666   & NGC 2685 & 8  & 55 & 34.7 & 58 & 44 & 4  & (R)SB0+ pec     & -19.21 &  876   &  14.6 & 61$^L$ & 13 -4$^2$ (U. Major S.\ Spur) \\  
  4862   & NGC 2782 & 9  & 14 & 5.1  & 40 & 6  & 49 & SAB(rs)a 	      & -20.99 &  2540  &  35.9 & 20$^e$ & 21 -0$^2$ (Leo) \\  
  5060   & NGC 2893 & 9  & 30 & 17.0 & 29 & 32 & 24 & (R)SB0/a 	      & -18.14 &  1699  &  24.1 & 24$^e$ & 21 -0$^2$ (Leo) \\  
  5253   & NGC 2985 & 9  & 50 & 22.2 & 72 & 16 & 43 & (R')SA(rs)ab    & -20.72 &  1325  &  21.1 & 38$^t$ & 12 -7$^2$ (U.\ Major) \\  
  5351   & NGC 3067 & 9  & 58 & 21.0 & 32 & 22 & 12 & SAB(s)ab?       & -19.35 &  1486  &  21.5 & 70$^L$ & 21 -12$^2$ (Leo) \\  
  5559   & NGC 3190 & 10 & 18 & 5.6  & 21 & 49 & 55 & SA(s)a pec sp   & -19.88 &  1308  &  18.6 & 67$^L$ & HCG 44$^4$ \\  
  5906   & NGC 3380 & 10 & 48 & 12.2 & 28 & 36 & 07 & (R')SBa?	      & -18.69 &  1601  &  23.1 & 29$^e$ & LGG 227$^1$ \\  
  5960   & NGC 3413 & 10 & 51 & 20.7 & 32 & 45 & 59 & S0 	      & -17.39 &   643  &  10.5 & 77$^t$ & -- \\  
  6001   & NGC 3442 & 10 & 53 & 8.1  & 33 & 54 & 37 & Sa? 	      & -17.77 &  1730  &  25.1 & 47$^L$ & 21 -9$^2$ (Leo) \\  
  6118   & NGC 3504 & 11 & 3  & 11.2 & 27 & 58 & 21 & (R)SAB(s)ab     & -20.28 &  1536  &  22.2 & 24$^e$ & 21 -7$^2$ (Leo) \\  
  6283   & NGC 3600 & 11 & 15 & 52.0 & 41 & 35 & 29 & Sa? 	      & -17.17 &   713  &  12.0 & 90$^e$ & 15 -9$^2$ (Leo Spur) \\  
  6621   & NGC 3786 & 11 & 39 & 42.5 & 31 & 54 & 33 & (R')SAB(r)a pec & -19.74 &  2742  &  38.7 & 58$^L$ & 13 -0$^2$ (U. Major S.\ Spur) \\  
  6623   & NGC 3788 & 11 & 39 & 44.6 & 31 & 55 & 52 & SAB(rs)ab pec   & -20.24 &  2670  &  37.7 & 74$^L$ & 13 -0$^2$ (U. Major S.\ Spur) \\  
  6742   & NGC 3870 & 11 & 45 & 56.6 & 50 & 12 & 00 & S0?             & -17.28 &   752  &  13.0 & 36$^L$ & 12 -1$^2$ (U.\ Major) \\  
  6786   & NGC 3900 & 11 & 49 & 9.4  & 27 & 1  & 19 & SA(r)0+ 	      & -19.94 &  1799  &  25.9 & 65$^t$ & 13 -9$^2$ (U. Major S.\ Spur) \\  
  6787   & NGC 3898 & 11 & 49 & 15.4 & 56 & 5  & 4  & SA(s)ab 	      & -20.25 &  1171  &  18.9 & 67$^t$ & 12 -3$^2$ (U.\ Major) \\  
  7166   & NGC 4151 & 12 & 10 & 32.6 & 39 & 24 & 21 & (R')SAB(rs)ab:  & -20.84 &   998  &  15.9 & 20$^t$ & 12 -6$^2$ (U.\ Major) \\  
  7256   & NGC 4203 & 12 & 15 & 5.0  & 33 & 11 & 50 & SAB0-: 	      & -19.47 &  1091  &  16.9 & 40$^t$ & 14 -1$^2$ (Coma--Sculptor) \\  
  7489   & NGC 4369 & 12 & 24 & 36.2 & 39 & 22 & 59 & (R)SA(rs)a      & -18.87 &  1027  &  16.4 & 17$^L$ & 12 -6$^2$ (U.\ Major) \\  
  7506   & NGC 4384 & 12 & 25 & 12.0 & 54 & 30 & 22 & Sa              & -19.70 &  2532  &  37.0 & 41$^L$ & 42 +11$^2$ (Canes Venatici) \\  
  7704   & NGC 4509 & 12 & 33 &  6.8 & 32 & 5  & 30 & Sab pec? 	      & -16.17 &  937   &  14.8 & 60$^e$ & LGG 279$^1$ \\  
  7989   & NGC 4725 & 12 & 50 & 26.6 & 25 & 30 & 3  & SAB(r)ab pec    & -21.69 &  1208  &  18.2 & 51$^t$ & 14 -2$^2$ (Coma--Sculptor) \\  
  8271   & NGC 5014 & 13 & 11 & 31.2 & 36 & 16 & 55 & Sa? sp 	      & -18.44 &  1128  &  17.7 & 60$^e$ & 43 -1$^2$ (Canes V.\ Spur) \\  
  8699   & NGC 5289 & 13 & 45 & 8.7  & 41 & 30 & 12 & (R)SABab:	      & -19.70 &  2521  &  36.7 & 72$^t$ & LGG 361$^1$ \\  
  8805   & NGC 5347 & 13 & 53 & 17.8 & 33 & 29 & 27 & (R')SB(rs)ab    & -19.69 &  2384  &  34.5 & 30$^e$ & 42 -0$^2$ (Canes Venatici) \\  
  8863   & NGC 5377 & 13 & 56 & 16.7 & 47 & 14 & 8  & (R)SB(s)a       & -20.32 &  1791  &  27.2 & 51$^t$ & 42 -0$^2$ (Canes Venatici) \\  
  9133   & NGC 5533 & 14 & 16 & 7.7  & 35 & 20 & 38 & SA(rs)ab 	      & -21.47 &  3861  &  54.3 & 53$^t$ & LGG 380$^1$ \\  
  9644   & --       & 14 & 59 & 34.3 & 27 & 6  & 58 & SB(r)a 	      & -19.25 &  6665  &  91.5 & 19$^e$ & -- \\  
  10448  & NGC 6186 & 16 & 34 & 25.5 & 21 & 32 & 27 & (R')SB(s)a      & -21.65 &  11351 & 153.8 & 30$^e$ & -- \\  
  11269  & NGC 6667 & 18 & 30 & 39.8 & 67 & 59 & 13 & SABab? pec      & -20.11 &  2581  &  38.3 & 56$^L$ & 70 -0$^2$ (--) \\  
  11670  & NGC 7013 & 21 & 3  & 33.6 & 29 & 53 & 51 & SA(r)0/a 	      & -19.53 &  775   &  12.7 & 68$^t$ & 65 +5$^2$ (Pegasus Spur) \\  
  11852  & --       & 21 & 55 & 59.3 & 27 & 53 & 54 & SBa? 	      & -20.25 &  5846  &  80.0 & 50$^t$ & -- \\  
  11914  & NGC 7217 & 22 & 7  & 52.4 & 31 & 21 & 33 & (R)SA(r)ab      & -20.45 &  949   &  14.9 & 31$^t$ & 65 +1$^2$ (Pegasus Spur) \\  
  11951  & NGC 7231 & 22 & 12 & 30.1 & 45 & 19 & 42 & SBa 	      & -19.14 &  1086  &  17.4 & 70$^L$ & -- \\  
  12043  & NGC 7286 & 22 & 27 & 50.5 & 29 & 5  & 45 & S0/a 	      & -17.15 &  1007  &  15.4 & 67$^t$ & -- \\  
  12276  & NGC 7440 & 22 & 58 & 32.5 & 35 & 48 & 9  & SB(r)a 	      & -20.59 &  5662  &  77.7 & 37$^L$ & -- \\  
  12713  & --       & 23 & 38 & 14.4 & 30 & 42 & 29 & S0/a 	      & -14.35 &   295  &   5.7 & 72$^t$ & 65 +4$^2$ (Pegasus Spur) \\  
  12815  & NGC 7771 & 23 & 51 & 24.9 & 20 & 6  & 43 & SB(s)a 	      & -21.60 &  4307  &  58.5 & 61$^L$ & LGG 483$^1$ \\ 
\end{longtable}}

\LTcapwidth=6.3in

\normalsize{
\begin{longtable}{rcr@{ $\times$ }r@{ $\times$ }rccll}

  \caption{Observational parameters: (1) UGC number, (2) observation
  date, (3) spatial resolution of the unsmoothed data, (4) velocity
  resolution before Hanning smoothing, (5) rms noise level in the
  unsmoothed channel maps, (6) 1 sigma column density in the \HI\
  maps shown in the atlas, (7) previous 21cm synthesis observations
  (references are explained at the bottom of the table) and (8)
  remarks. \label{table:observations}}  \\   
  
  \hline \hline
  \multicolumn{1}{c}{UGC} & observation date &
  \multicolumn{3}{c}{resolution} & $\sigma_{\mathrm{ch}}$ & 
  $\sigma_{\mathrm{map}}$ & previous & remarks \\    

  & & RA & Dec & v \hspace{0.22cm}  & & & observations & \\ 
  
  & & \multicolumn{1}{c}{\arcsec} & \multicolumn{1}{l}{\arcsec} & 
  km~s$^{-1}$\hspace{-0.2cm} & mJy/beam & $10^{19}$ atoms~cm$^{-2}$ &
  & \\  
  
  (1) & (2) & \multicolumn{2}{c}{(3)} & (4)\hspace{0.2cm} & (5) & (6)
  & \hspace{0.35cm}(7) & \hspace{0.3cm} (8) \\   
  \hline
  \endfirsthead  

  \caption[]{observational parameters: continued}\\  

  \hline \hline
  \multicolumn{1}{c}{UGC} & observation date &
  \multicolumn{3}{c}{resolution} & $\sigma_{\mathrm{ch}}$ & 
  $\sigma_{\mathrm{map}}$ & previous & remarks \\    

  & & RA & Dec & v \hspace{0.22cm}  & & & observations & \\  
  
  & & \multicolumn{1}{c}{\arcsec} & \multicolumn{1}{l}{\arcsec} & 
  km~s$^{-1}$\hspace{-0.2cm} & mJy/beam & $10^{19}$
  atoms~cm$^{-2}$ & & \\ 
  
  (1) & (2) & \multicolumn{2}{c}{(3)} & (4)\hspace{0.2cm} & (5) & (6)
  & \hspace{0.35cm}(7) & \hspace{0.3cm} (8) \\   
  \hline
  \endhead  

  \hline
  \multicolumn{9}{l}{$^\dagger$ observed before the upgrade of
  the WSRT.} 
  \endfoot

  \hline
  \multicolumn{9}{l}{$^\dagger$ observed before the upgrade of
  the WSRT.} \\[0.25cm]
  \multicolumn{9}{l}{explanation of the references in column (7):} \\[0.1cm]
  A79  & \multicolumn{4}{l}{\citet{Allsopp79}}     & \multicolumn{1}{r}{S87}  & \multicolumn{3}{l}{\citet{Simkin87}}      \\
  B77  & \multicolumn{4}{l}{\citet{Bosma77}}       & \multicolumn{1}{r}{S94}  & \multicolumn{3}{l}{\citet{Smith94}}       \\
  B92  & \multicolumn{4}{l}{\citet{Broeils92}}     & \multicolumn{1}{r}{VD88} & \multicolumn{3}{l}{\citet{VanDriel88}}    \\
  B94  & \multicolumn{4}{l}{\citet{Broeils94}}     & \multicolumn{1}{r}{VD89} & \multicolumn{3}{l}{\citet{VanDriel89}}    \\
  C89  & \multicolumn{4}{l}{\citet{Carilli89}}     & \multicolumn{1}{r}{VD91} & \multicolumn{3}{l}{\citet{VanDriel91}}    \\
  C92  & \multicolumn{4}{l}{\citet{Carilli92}}     & \multicolumn{1}{r}{VD94} & \multicolumn{3}{l}{\citet{VanDriel94}}    \\ 
  GR02 & \multicolumn{4}{l}{\citet{Garcia-Ruiz02}} & \multicolumn{1}{r}{VM83a}& \multicolumn{3}{l}{\citet{VanMoorsel83}}  \\
  H82  & \multicolumn{4}{l}{\citet{Heckman82}}     & \multicolumn{1}{r}{VM83b}& \multicolumn{3}{l}{\citet{VanMoorsel83b}} \\
  H00  & \multicolumn{4}{l}{\citet{Haynes00}}      & \multicolumn{1}{r}{VM88} & \multicolumn{3}{l}{\citet{VanMoorsel88}}  \\
  K84  & \multicolumn{4}{l}{\citet{Knapp84}}       & \multicolumn{1}{r}{VM95} & \multicolumn{3}{l}{\citet{Verdes-Montenegro95}} \\
  M99  & \multicolumn{4}{l}{\citet{Mundell99}}     & \multicolumn{1}{r}{VW83} & \multicolumn{3}{l}{\citet{VanWoerden83}}  \\
  O93  & \multicolumn{4}{l}{\citet{Oosterloo93}}   & \multicolumn{1}{r}{W84}  & \multicolumn{3}{l}{\citet{Wevers84}}      \\
  P92  & \multicolumn{4}{l}{\citet{Pedlar92}}      & \multicolumn{1}{r}{W88}  & \multicolumn{3}{l}{\citet{Warmels88b}}    \\
  S80  & \multicolumn{4}{l}{\citet{Shane80}}       & \multicolumn{1}{r}{W91}  & \multicolumn{3}{l}{\citet{Williams91}}    \\
  S84  & \multicolumn{4}{l}{\citet{Sancisi84}}
  \endlastfoot

  89    & Sep. 2000 	      & 16.1  & 37.9  & 10.03 & 0.64 & 3.25 & O93      & \rdelim\}{2}{2cm}[1 pointing] \\
  94    & Sep. 2000	      & 16.1  & 37.9  & 10.03 & 0.64 & 2.86 & O93      & \\      
  232   & July 2000 	      & 15.7  & 29.9  & 10.04 & 0.78 & 4.97 & --       & \\
  499   & Aug. 2000 	      & 11.1  & 19.4  & 10.04 & 0.69 & 2.56 & H82, S87 & \rdelim\}{2}{2cm}[1 pointing] \\
  508   & Aug. 2000 	      & 11.1  & 19.4  & 10.04 & 0.69 & 2.82 & --       & \\
  624   & July 2000 	      & 10.7  & 24.2  & 10.04 & 0.61 & 11.3 & --       & \\
  798   & Aug. 2000 	      & 12.6  & 23.2  & 10.04 & 0.73 & 3.13 & --       & \\
  1310 	& June 2000 	      & 12.0  & 31.8  & 9.98  & 0.90 & 3.66 & VM88     & \\
  1541 	& Aug. 2000 	      & 16.1  & 27.0  & 10.07 & 0.79 & 7.06 & --       & \\
  2045 	& Aug. 2000  	      & 15.9  & 34.0  & 9.94  & 0.67 & 2.75 & --       & \\
  2141 	& Jan. 1994$^\dagger$ & 8.3   & 19.4  & 4.98  & 3.51 & 71.8 & --       & \\
  2154 	& July 1997$^\dagger$ & 10.1  & 18.0  & 19.80 & 1.20 & 2.88 & A79, S84 & \\
  2183 	& Apr. 2000 	      & 10.6  & 26.0  & 9.94  & 0.79 & 6.52 & --       & \\
  2487 	& Dec. 2000 	      & 11.8  & 22.2  & 10.06 & 0.87 & 1.29 & --       & \\
  2916 	& Apr. 2000 	      & 10.8  & 13.2  & 10.03 & 0.59 & 16.8 & --       & \\
  2941 	& July 2000 	      & 10.7  & 32.3  & 5.06  & 1.31 & 6.47 & --       & \\
  2953 	& Sep. 2001 	      & 10.7  & 14.4  & 9.91  & 0.60 & 1.39 & --       & \\
  3205  & Dec. 2000 	      & 10.6  & 24.5  & 10.01 & 0.61 & 15.5 & --       & \\
  3354  & Nov. 2000 	      & 16.9  & 21.1  & 9.98  & 0.57 & 5.93 & --       & \\
  3382  & Dec. 2000 	      & 12.2  & 14.1  & 5.04  & 0.75 & 1.20 & --       & \\
  3407  & Dec. 2000 	      & 12.7  & 17.8  & 10.01 & 0.71 & 2.38 & --       & \\
  3426  & Jan. 2002 	      & 12.3  & 13.0  & 5.03  & 0.54 & 1.20 & --       & \\
  3546  & Dec. 2000 	      & 12.2  & 14.9  & 9.95  & 0.63 & 1.13 & VW83     & \\
  3580  & Jan. 1994$^\dagger$ & 9.1   & 10.2  & 4.98  & 2.50 & 5.45 & B94      & \\
  3642  & May 2000  	      & 12.6  & 13.4  & 10.03 & 0.75 & 3.69 & --       & \\
  3965  & Apr. 2000 	      & 11.7  & 21.4  & 2.51  & 1.84 & 1.95 & --       & \\
  3993  & July 2000 	      & 12.2  & 12.3  & 5.03  & 0.82 & 1.32 & --       & \\
  4458  & July 2002 	      & 15.5  & 43.6  & 10.04 & 0.74 & 3.14 & --       & \\
  4605  & Apr. 2000 	      & 12.8  & 13.7  & 9.94  & 0.69 & 18.4 & W88      & \\
  4637  & July 2002 	      & 12.6  & 12.4  & 9.94  & 0.68 & 0.74 & --       & \\
  4666  & Aug. 2000 	      & 11.8  & 15.6  & 9.91  & 0.69 & 2.95 & S80      & \\
  4862  & July 2000 	      & 15.5  & 24.8  & 5.00  & 0.99 & 6.07 & S94      & \\
  5060  & July 2000 	      & 15.6  & 33.2  & 4.99  & 0.88 & 2.04 & --       & \\
  5253  & Aug. 1993$^\dagger$ & 11.6  & 12.4  & 19.92 & 1.82 & 3.64 & O93      & \\
  5351  & Aug. 2000 	      & 14.0  & 20.8  & 9.94  & 0.75 & 4.11 & C89, C92 & \\
  5559  & June 1996$^\dagger$ & 9.8   & 26.3  & 19.92 & 1.98 & 31.1 & W91      & \\
  5906  & Aug. 2000 	      & 34.6  & 15.8  & 4.99  & 1.05 & 2.04 & --       & \\
  5960  & Apr. 2000 	      & 21.7  & 11.6  & 4.97  & 1.40 & 6.45 & --       & \\
  6001  & Aug. 2000 	      & 21.4  & 11.7  & 4.99  & 0.97 & 5.07 & --       & \\
  6118  & July 2000 	      & 35.2  & 15.7  & 9.94  & 0.78 & 4.35 & VM83b    & \\
  6283  & Apr. 1996$^\dagger$ & 18.7  & 12.5  & 4.97  & 3.11 & 28.9 & GR02     & \\
  6621  & July 2002 	      & 31.2  & 15.9  & 9.97  & 0.59 & 3.70 & O93      & \rdelim\}{2}{2cm}[1 pointing] \\
  6623  & July 2002 	      & 31.2  & 15.9  & 9.97  & 0.59 & 3.70 & O93      & \\
  6742  & Apr. 2000 	      & 14.1  & 12.8  & 4.97  & 0.81 & 1.84 & --       & \\
  6786  & Aug. 2000 	      & 15.5  & 36.0  & 9.95  & 0.74 & 1.48 & VD89, H00& \\
  6787  & Apr. 2000 	      & 12.6  & 13.8  & 9.92  & 0.65 & 2.14 & VD94     & \\
  7166  & June 1997$^\dagger$ & 20.8  & 12.9  & 4.98  & 2.98 & 1.73 & B77, P92, M99& \\
  7256  & July 2000 	      & 30.7  & 15.3  & 4.98  & 1.76 & 2.29 & VD88     & \\
  7489  & Aug. 1997$^\dagger$ & 18.2  & 11.4  & 4.98  & 4.30 & 5.18 & --       & \\
  7506  & Aug. 2000 	      & 13.0  & 10.3  & 5.00  & 0.93 & 13.1 & --       & \\
  7704  & Aug. 2000 	      & 21.3  & 10.0  & 4.98  & 0.93 & 4.69 & --       & \\
  7989  & Oct. 1994$^\dagger$ & 8.0   & 13.2  & 19.80 & 1.13 & 9.73 & W84      & \\
  8271  & July 2000 	      & 10.2  & 19.9  & 4.98  & 0.91 & 0.98 & --       & \\
  8699  & Aug. 2000 	      & 16.9  & 10.7  & 9.97  & 0.68 & 12.4 & VM83a    & \\
  8805  & July 1997$^\dagger$ & 19.5  & 10.6  & 5.00  & 4.41 & 2.49 & --       & \\
  8863  & Aug. 2000 	      & 15.8  & 22.4  & 9.95  & 0.66 & 2.11 & --       & \\
  9133  & July 2000 	      & 15.8  & 28.8  & 10.02 & 0.70 & 2.44 & B92, B94 & \\
  9644  & July 2000 	      & 10.5  & 25.6  & 5.08  & 1.08 & 9.41 & --       & \\
  10448 & July 2000 	      & 10.6  & 32.1  & 5.15  & 1.10 & 5.26 & --       & \\
  11269 & July 2000 	      & 12.2  & 14.4  & 9.97  & 0.59 & 1.78 & --       & \\
  11670 & May 2000 	      & 11.3  & 25.0  & 9.91  & 0.73 & 2.37 & K84      & \\
  11852 & July 2000 	      & 16.1  & 35.7  & 10.08 & 0.78 & 4.12 & --       & \\
  11914 & Sep. 2000 	      & 10.5  & 23.5  & 9.92  & 0.68 & 2.25 & VM95     & \\
  11951 & Jan. 1994$^\dagger$ & 10.1  & 14.1  & 4.98  & 3.04 & 54.5 & --       & \\
  12043 & Apr. 2000 	      & 9.9   & 25.9  & 4.98  & 1.22 & 9.50 & --       & \\
  12276 & July 2002 	      & 10.4  & 19.3  & 5.05  & 0.84 & 9.45 & --       & \\
  12713 & June 2000 	      & 11.6  & 22.0  & 4.97  & 0.95 & 1.78 & VD91     & \\
  12815 & Aug. 2000 	      & 10.4  & 33.4  & 10.03 & 0.70 & 9.62 & --       & \\
\end{longtable}}

\LTcapwidth=7.in

\small{
\begin{longtable}{rrrrrr@{\hspace{0.35cm}}rrrrrcccc}

  \caption{\HI\ properties: (1) UGC number, (2) systemic velocity, (3) 
  distance, (4) line width of global profile at the 20\% level, (5)
  idem at the 50\% level, (6) total \HI\ flux, (7) total \HI\ mass,
  (8) \HI\ radius in \arcsec, (9) idem in kpc, (10) average face-on
  \HI\ surface density within \HI\ radius, (11) idem within optical
  radius, (12) amount of asymmetry in global profile, (13)
  morphological asymmetry, (14) kinematical asymmetry, and (15)
  evidence for current interaction/merging/accretion.  
  \label{table:data}}\\  

  \hline \hline
  \multicolumn{1}{c}{UGC} & \multicolumn{1}{c}{$V_{\mathrm{sys}}$} & 
  \multicolumn{1}{c}{$D$} & \multicolumn{1}{c}{$W_{20}^c$} &
  \multicolumn{1}{c}{$W_{50}^c$} & \multicolumn{1}{c}{$\int
  F\,\mathrm{dv}$} & \multicolumn{1}{c}{$M_{\HI}$} &
  \multicolumn{1}{c}{${\mathrm R}_\HI$} & \multicolumn{1}{c}{${\mathrm
  R}_\HI$} & \multicolumn{1}{c}{$<\!\Sigma_\HI\!>_{{\mathrm R}_\HI}$}
  & \multicolumn{1}{c}{$<\!\Sigma_\HI\!>_{{\mathrm R}_{25}}$} &
  \multicolumn{4}{c}{Asymmetries} \\   
  
  & \multicolumn{1}{c}{km~s$^{-1}$} & \multicolumn{1}{c}{Mpc} &
  \multicolumn{1}{c}{km~s$^{-1}$} & \multicolumn{1}{c}{km~s$^{-1}$} &
  \multicolumn{1}{c}{Jy~km~s$^{-1}$} & \multicolumn{1}{c}{$10^9
  \msun$} & \multicolumn{1}{c}{\arcsec} & \multicolumn{1}{c}{kpc} &
  \multicolumn{1}{c}{\msunpc2} & \multicolumn{1}{c}{\msunpc2} &
  prof. & morph. & kin. & interac. \\   

  \multicolumn{1}{c}{(1)} & \multicolumn{1}{c}{(2)} &
  \multicolumn{1}{c}{(3)} & \multicolumn{1}{c}{(4)} &
  \multicolumn{1}{c}{(5)} & \multicolumn{1}{c}{(6)} &
  \multicolumn{1}{c}{(7)} & \multicolumn{1}{c}{(8)} & 
  \multicolumn{1}{c}{(9)} & \multicolumn{1}{c}{(10)} & 
  \multicolumn{1}{c}{(11)} & \multicolumn{1}{c}{(12)} & 
  \multicolumn{1}{c}{(13)} & \multicolumn{1}{c}{(14)} &
  \multicolumn{1}{c}{(15)} \\  
  \hline  
  \endfirsthead  

  \caption[]{\HI\ properties: continued}\\
 
  \hline \hline
  \multicolumn{1}{c}{UGC} & \multicolumn{1}{c}{$V_{\mathrm{sys}}$} &
  \multicolumn{1}{c}{$D$} & \multicolumn{1}{c}{$W_{20}^c$} &
  \multicolumn{1}{c}{$W_{50}^c$} & \multicolumn{1}{c}{$\int
  F\,\mathrm{dv}$} & \multicolumn{1}{c}{$M_{\HI}$} &
  \multicolumn{1}{c}{${\mathrm R}_\HI$} & \multicolumn{1}{c}{${\mathrm
  R}_\HI$} & \multicolumn{1}{c}{$<\!\Sigma_\HI\!>_{{\mathrm R}_\HI}$}
  & \multicolumn{1}{c}{$<\!\Sigma_\HI\!>_{{\mathrm R}_{25}}$} &  
  \multicolumn{4}{c}{Asymmetries} \\  
  
  & \multicolumn{1}{c}{km~s$^{-1}$} & \multicolumn{1}{c}{Mpc} &
  \multicolumn{1}{c}{km~s$^{-1}$} & \multicolumn{1}{c}{km~s$^{-1}$} &
  \multicolumn{1}{c}{Jy~km~s$^{-1}$} & \multicolumn{1}{c}{$10^9
  \msun$} & \multicolumn{1}{c}{\arcsec} & \multicolumn{1}{c}{kpc} &
  \multicolumn{1}{c}{\msunpc2} & \multicolumn{1}{c}{\msunpc2} &
  prof. & morph. & kin. & interac. \\   

  \multicolumn{1}{c}{(1)} & \multicolumn{1}{c}{(2)} &
  \multicolumn{1}{c}{(3)} & \multicolumn{1}{c}{(4)} &
  \multicolumn{1}{c}{(5)} & \multicolumn{1}{c}{(6)} &
  \multicolumn{1}{c}{(7)} & \multicolumn{1}{c}{(8)} & 
  \multicolumn{1}{c}{(9)} & \multicolumn{1}{c}{(10)} & 
  \multicolumn{1}{c}{(11)} & \multicolumn{1}{c}{(12)} & 
  \multicolumn{1}{c}{(13)} & \multicolumn{1}{c}{(14)} &
  \multicolumn{1}{c}{(15)} \\  
  \hline  
  \endhead  

  \hline 
  \multicolumn{15}{l}{$^\#$ Average surface density $ < 1 \, \msunpc2$
  everywhere.} \\
  \multicolumn{15}{l}{$^\dagger$ Total flux and \HI\ mass measured
  from $30\arcsec$ data.} \\ 
  \endfoot

  \hline 
  \multicolumn{15}{l}{$^\$$ Includes emission from UGC~12813.}
  \endlastfoot

  89     & 4555  &  62.1 & 440  & 386  & 9.44   	  & 8.63    	   & 77		& 24	  & 3.5    \hspace{0.3cm} & 5.1\hspace{0.3cm} & *  	& *	& 	& +   	\\
  94     & 4589  &  62.6 & 319  & 300  & 12.19  	  & 11.28    	   & 85 	& 26 	  & 4.0    \hspace{0.3cm} & 7.2\hspace{0.3cm} & *  	& *	& 	& +   	\\
  232    & 4839  &  66.3 & 276  & 236  & 7.66   	  & 7.95    	   & 82 	& 27 	  & 2.1    \hspace{0.3cm} & 3.3\hspace{0.3cm} &   	& *	& 	&    	\\
  499    & 4534  &  62.0 &  91  &  65  & 23.9   	  & 21.45    	   & 104 	& 31 	  & 1.6    \hspace{0.3cm} & 1.8\hspace{0.3cm} &   	& **	& n.a.	& +   	\\
  508    & 4647  &  63.6 & 478  & 453  & 7.71   	  & 7.40    	   & 94 	& 29 	  & 2.0    \hspace{0.3cm} & 2.0\hspace{0.3cm} &   	& *	& 	& +   	\\
  624    & 4772  &  65.1 & 560  & 518  & 14.39  	  & 14.42   	   & 92 	& 29 	  & 4.3    \hspace{0.3cm} & 7.7\hspace{0.3cm} &   	& **	& ***	& +   	\\
  798    & 4896  &  66.7 & 220  & 205  & 3.88   	  & 4.08    	   & 72 	& 23 	  & 1.8    \hspace{0.3cm} & 2.4\hspace{0.3cm} &   	& 	& 	&    	\\
  1310   & 2958  &  40.2 & 186  & 117  & 6.16   	  & 2.33    	   & 50 	& 9.7	  & 4.8    \hspace{0.3cm} &14.3\hspace{0.3cm} & **      & n.a.	& n.a.	& +   	\\
  1541   & 5649  &  77.0 & 446  & 421  & 7.94   	  & 11.14    	   & 75 	& 28 	  & 2.8    \hspace{0.3cm} & 3.9\hspace{0.3cm} & ***     & *	& *	&    	\\
  2045   & 1527  &  21.4 & 334  & 289  & 18.89  	  & 2.09    	   & 94 	& 9.9 	  & 4.2    \hspace{0.3cm} & 3.6\hspace{0.3cm} &    	&  	& 	&    	\\
  2141   & 987   &  14.3 & 232  & 201  & 47.94  	  & 2.30    	   & 155 	& 10	  & 4.5    \hspace{0.3cm} &11.3\hspace{0.3cm} &    	& *	& 	&    	\\
  2154   & 695   &  10.9 & 529  & 497  & 80.13  	  & 2.26    	   & --$^\#$	&--$^\#$  & --$^\#$\hspace{0.3cm} & 0.7\hspace{0.3cm} & n.a.    & n.a.	& n.a. 	& +  	\\
  2183   & 1540  &  21.5 & 292  & 268  & 32.99  	  & 3.63   	   & 164 	& 17 	  & 3.0    \hspace{0.3cm} & 6.8\hspace{0.3cm} & *  	& *	& *	&    	\\
  2487   & 4950  &  67.4 & 466  & 437  & 15.93  	  & 17.09    	   & 120 	& 39 	  & 1.6    \hspace{0.3cm} & 1.8\hspace{0.3cm} &   	& 	& 	&    	\\
  2916   & 4518  &  63.5 & 357  & 337  & 23.08  	  & 21.94    	   & 108 	& 34 	  & 4.3    \hspace{0.3cm} & 7.3\hspace{0.3cm} & *  	& 	& **	& +   	\\
  2941   & 6261  &  83.9 & 160  & 139  & 9.84   	  & 16.32    	   & 75 	& 30 	  & 3.4    \hspace{0.3cm} & 5.4\hspace{0.3cm} &   	& *	& *	& +   	\\
  2953   & 894   &  15.1 & 484  & 464  & 119.73 	  & 6.43    	   & 300 	& 22 	  & 1.9    \hspace{0.3cm} & 2.2\hspace{0.3cm} &    	& 	& 	&    	\\
  3205   & 3587  &  48.7 & 437  & 419  & 16.38  	  & 9.18    	   & 118 	& 28 	  & 3.4    \hspace{0.3cm} & 4.7\hspace{0.3cm} &   	& 	& 	&    	\\
  3354   & 3084  &  43.6 & 423  & 398  & 17.71  	  & 7.95    	   & 77 	& 16 	  & 2.0    \hspace{0.3cm} & 2.3\hspace{0.3cm} & *       & 	& 	&    	\\
  3382   & 4501  &  62.8 & 204  & 193  & 5.68   	  & 5.27    	   & 78 	& 24 	  & 2.0    \hspace{0.3cm} & 2.8\hspace{0.3cm} &   	& 	& 	&    	\\
  3407   & 3606  &  49.8 & 316  & 298  & 3.13   	  & 1.82   	   & 50 	& 12 	  & 2.6    \hspace{0.3cm} & 3.1\hspace{0.3cm} & *       & 	& 	& +   	\\
  3426   & 4005  &  56.6 & 161  & 114  & 3.27   	  & 2.53    	   & --$^\#$ 	&--$^\#$  & --$^\#$\hspace{0.3cm} & 0.2\hspace{0.3cm} & *       & n.a.	& n.a.	& +   	\\
  3546   & 1838  &  27.3 & 367  & 347  & 13.74  	  & 2.42    	   & 132 	& 18 	  & 2.0    \hspace{0.3cm} & 2.2\hspace{0.3cm} &   	& 	& 	&    	\\
  3580   & 1200  &  19.2 & 241  & 224  & 43.24  	  & 3.76    	   & 199 	& 18 	  & 3.0    \hspace{0.3cm} & 5.4\hspace{0.3cm} & *  	& 	& * 	&    	\\
  3642   & 4498  &  62.9 & 469  & 441  & 38.13  	  & 35.62    	   & 171 	& 52 	  & 3.1    \hspace{0.3cm} & 4.7\hspace{0.3cm} & *       &  	& 	&    	\\
  3965   & 4588  &  62.5 &  93  &  74  & 15.67  	  & 14.47    	   & 106 	& 33 	  & 3.2    \hspace{0.3cm} & 5.3\hspace{0.3cm} &   	&  	& 	&    	\\
  3993   & 4366  &  61.9 & 209  & 192  & 7.73   	  & 6.99    	   & 82 	& 24 	  & 1.6    \hspace{0.3cm} & 1.7\hspace{0.3cm} &   	& *	& 	&    	\\
  4458   & 4757  &  64.2 & 284  & 245  & 12.24  	  & 11.82   	   & 110 	& 34 	  & 2.0    \hspace{0.3cm} & 3.1\hspace{0.3cm} &   	& **	& 	& +   	\\
  4605   & 1350  &  20.9 & 432  & 392  & 58.03  	  & 5.99    	   & 277 	& 28 	  & 1.8    \hspace{0.3cm} & 3.0\hspace{0.3cm} &   	& 	& 	&    	\\
  4637   & 1415  &  22.4 & 399  & 243  & 22.34  	  & 2.62    	   & --$^\#$ 	&--$^\#$  & --$^\#$\hspace{0.3cm} & 0.9\hspace{0.3cm} & **      & ***	& n.a.	& +   	\\
  4666   & 876   &  14.6 & 303  & 284  & 29.07  	  & 1.47    	   & 184 	& 12 	  & 2.2    \hspace{0.3cm} & 2.5\hspace{0.3cm} & *       & *	& 	&    	\\
  4862   & 2540  &  35.9 & 192  & 156  & 16.42  	  & 5.07    	   & 90 	& 16 	  & 3.1    \hspace{0.3cm} & 2.2\hspace{0.3cm} & ***     & ***	& n.a.	& +   	\\
  5060   & 1699  &  24.1 & 185  &  86  & 5.13   	  & 0.70   	   & 60		& 7.0	  & 2.3    \hspace{0.3cm} & 3.7\hspace{0.3cm} & **      & ***	& 	&    	\\
  5253   & 1325  &  21.1 & 320  & 296  & 132.4  	  & 13.92          & 330 	& 34 	  & 2.1    \hspace{0.3cm} & 3.8\hspace{0.3cm} & *       & ***	& 	& +   	\\
  5351   & 1486  &  21.5 & 276  & 187  & 9.23   	  & 0.99    	   & 61 	& 6.3 	  & 3.4    \hspace{0.3cm} & 3.1\hspace{0.3cm} & **      & **	& **	& +   	\\
  5559   & 1308  &  18.6 & 473  & 449  & 7.32   	  & 0.60    	   & 88 	& 7.7 	  & 2.6    \hspace{0.3cm} & 1.5\hspace{0.3cm} &   	& n.a.	& n.a.	& +   	\\
  5906   & 1601  &  23.1 & 123  & 110  & 2.53   	  & 0.31    	   & 47 	& 5.3 	  & 2.0    \hspace{0.3cm} & 1.9\hspace{0.3cm} &   	& *	& 	&    	\\
  5960   &  643  &  10.5 & 180  & 154  & 28.42  	  & 0.75    	   & 109	& 5.7 	  & 6.4    \hspace{0.3cm} &15.9\hspace{0.3cm} & **      & *	& **	&    	\\
  6001   & 1730  &  25.1 & 153  & 121  & 3.58   	  & 0.33    	   & 44 	& 5.3	  & 4.2    \hspace{0.3cm} & 9.6\hspace{0.3cm} & *       & n.a.	& n.a.	&    	\\
  6118   & 1536  &  22.2 & 219  & 194  & 7.20   	  & 0.84    	   & 89 	& 9.5 	  & 2.7    \hspace{0.3cm} & 3.0\hspace{0.3cm} & *  	& 	& 	&    	\\
  6283   &  713  &  12.0 & 216  & 200  & 57.38  	  & 2.00    	   & 204 	& 12 	  & 3.1    \hspace{0.3cm} & 5.5\hspace{0.3cm} &   	& *	& 	&    	\\
  6621   & 2742  &  38.7 & 420  & 375  & 3.69$^\dagger$   & 1.30$^\dagger$ & 57 	& 10 	  & 4.1    \hspace{0.3cm} & 4.1\hspace{0.3cm} &    	& ***	& n.a.	& +   	\\
  6623   & 2670  &  37.7 & 520  & 438  & 11.51$^\dagger$  & 3.95$^\dagger$ & 79 	& 15 	  & 3.5    \hspace{0.3cm} & 4.4\hspace{0.3cm} & ***     & ***	& n.a.	& +   	\\
  6742   &  752  &  13.0 & 127  &  84  & 4.08   	  & 0.17    	   & 51 	& 3.2 	  & 4.3    \hspace{0.3cm} & 8.0\hspace{0.3cm} & *  	& **	& 	&    	\\
  6786   & 1799  &  25.9 & 445  & 423  & 24.91  	  & 3.93    	   & 143 	& 18 	  & 2.4    \hspace{0.3cm} & 3.4\hspace{0.3cm} &   	& 	& 	&    	\\
  6787   & 1171  &  18.9 & 484  & 466  & 46.82  	  & 3.96    	   & 256 	& 23 	  & 1.7    \hspace{0.3cm} & 2.5\hspace{0.3cm} & * 	& *	& 	& +   	\\
  7166   &  998  &  15.9 & 140  & 121  & 67.58  	  & 4.05    	   & 304 	& 24 	  & 2.1    \hspace{0.3cm} & 2.2\hspace{0.3cm} & **      & *	& *	&    	\\
  7256   & 1091  &  16.9 & 271  & 239  & 48.11  	  & 3.21    	   & 226 	& 19 	  & 1.3    \hspace{0.3cm} & 1.8\hspace{0.3cm} & *  	& ** 	& n.a.	& +   	\\
  7489   & 1027  &  16.4 &  86  &  68  & 4.07  	  	  & 0.26    	   & 55		& 4.4 	  & 3.4    \hspace{0.3cm} & 2.7\hspace{0.3cm} &   	& n.a.	& n.a. 	&    	\\
  7506   & 2532  &  37.0 & 171  & 116  & 3.66   	  & 1.17    	   & 37 	& 6.7 	  & 4.1    \hspace{0.3cm} & 3.7\hspace{0.3cm} & *       & n.a.	& n.a.	&    	\\
  7704   & 937   &  14.8 &  93  &  61  & 6.49   	  & 0.34    	   & 58 	& 4.3 	  & 3.7    \hspace{0.3cm} & 9.4\hspace{0.3cm} &   	& * 	& 	&    	\\
  7989   & 1208  &  18.2 & 417  & 390  & 125.78 	  & 9.78    	   & 380 	& 34	  & 2.4    \hspace{0.3cm} & 2.7\hspace{0.3cm} & ***     & **	& *	& +   	\\
  8271   & 1128  &  17.7 & 166  &  76  & 10.37  	  & 0.78    	   & 89 	& 7.7 	  & 2.1    \hspace{0.3cm} & 3.7\hspace{0.3cm} & *  	& ***	& n.a.	& +   	\\
  8699   & 2521  &  36.7 & 387  & 368  & 10.22  	  & 3.26    	   & 100 	& 17 	  & 2.9    \hspace{0.3cm} & 4.6\hspace{0.3cm} &   	& 	& 	&    	\\
  8805   & 2384  &  34.5 & 129  &  82  & 15.45  	  & 4.34    	   & 137 	& 23 	  & 1.8    \hspace{0.3cm} & 3.0\hspace{0.3cm} &    	&  	& 	&    	\\
  8863   & 1791  &  27.2 & 391  & 367  & 14.44  	  & 2.52    	   & 148 	& 19 	  & 1.5    \hspace{0.3cm} & 1.6\hspace{0.3cm} & *  	& 	& *	&    	\\
  9133   & 3861  &  54.3 & 451  & 416  & 43.25  	  & 30.23    	   & 218 	& 57 	  & 1.9    \hspace{0.3cm} & 4.0\hspace{0.3cm} & **      & **	&  	&    	\\
  9644   & 6665  &  91.5 & 131  & 119  & 3.63   	  & 7.16    	   & 53		& 24 	  & 3.2    \hspace{0.3cm} & 4.4\hspace{0.3cm} &   	& 	& 	&    	\\
  10448  & 11351 & 153.8 &  99  &  85  & 1.76   	  & 9.82     	   & 40 	& 29 	  & 2.6    \hspace{0.3cm} & 2.0\hspace{0.3cm} &    	& 	& 	&    	\\
  11269  & 2581  &  38.3 & 415  & 344  & 31.32  	  & 10.86    	   & 161 	& 30 	  & 3.0    \hspace{0.3cm} & 6.1\hspace{0.3cm} & *  	& ***	& **	& +   	\\
  11670  & 775   &  12.7 & 342  & 325  & 28.05  	  & 1.08    	   & 158 	& 9.9 	  & 2.5    \hspace{0.3cm} & 2.5\hspace{0.3cm} & **      & 	& *	&    	\\
  11852  & 5846  &  80.0 & 329  & 305  & 19.55  	  & 29.6     	   & 137 	& 53 	  & 2.5    \hspace{0.3cm} & 4.2\hspace{0.3cm} & ** 	& 	& **	&    	\\ 
  11914  & 949   &  14.9 & 322  & 309  & 13.16  	  & 0.69   	   & 113 	& 8.1 	  & 3.0    \hspace{0.3cm} & 2.8\hspace{0.3cm} &   	& 	& 	&    	\\
  11951  & 1086  &  17.4 & 222  & 201  & 25.93  	  & 1.85    	   & 127 	& 10.7 	  & 4.4    \hspace{0.3cm} & 9.6\hspace{0.3cm} &   	& 	& 	& +   	\\
  12043  & 1007  &  15.4 & 190  & 176  & 20.96  	  & 1.18    	   & 143	& 10.8 	  & 2.4    \hspace{0.3cm} & 6.1\hspace{0.3cm} &   	& 	& 	&    	\\
  12276  & 5662  &  77.7 & 135  & 108  & 4.09   	  & 5.83    	   & 67 	& 26 	  & 2.2    \hspace{0.3cm} & 2.9\hspace{0.3cm} & *       & * 	& 	& +   	\\
  12713  &  295  &   5.7 & 147  & 101  & 10.79  	  & 0.08  	   & 84 	& 2.3 	  & 2.0    \hspace{0.3cm} & 4.3\hspace{0.3cm} & *  	& *	& *	&    	\\
  12815  & 4307  &  58.5 & 656  & 457  & 11.74$^\$$ 	  & 9.36$^\$$      & 103 	& 30 	  & 1.9    \hspace{0.3cm} & 2.6\hspace{0.3cm} & n.a.    & n.a.	& n.a.	& +   	\\
\end{longtable}}

\clearpage
\twocolumn


\appendix
\section{Atlas of \HI\ observations}
\label{app:atlas}
On the following pages, we present the \HI\ observations for all our 
sample galaxies. For each galaxy, we show a figure consisting of 6
panels: \newline
{\bf Upper left:} Grayscale and contour image of the integrated \HI\ 
distribution. Thin contours are at $2,4,8,16,\ldots \cdot
\sigma_{\mathrm {map}}$, with $\sigma_{\mathrm {map}}$ determined as
in Sect.~\ref{subsec:HImaps}; its value in atoms cm$^{-2}$ is given
in Table~\ref{table:observations}. The thick contour indicates the
1~\msunpc2\ level, corrected to face-on. For edge-on
galaxies, the correction to face-on could not be done, and we only
show the $\sigma_{\mathrm {map}}$-levels. The beam size for the data
shown here is given in the lower left. Note that this is not always
the same as the resolution given in Table~\ref{table:observations}, as
we sometimes needed to smooth the data in order to increase the
signal-to-noise ratio. \newline   
{\bf Lower left:} The same image overlayed on an optical R-band 
image. We show only the $2,8,32,128,\ldots \cdot \sigma_{\mathrm
{map}}$ and  1~\msunpc2\ contours here, and omit
the contours at $4,16,64,\ldots \cdot \sigma_{\mathrm {map}}$. The
images were taken as part of a  parallel project to image our sample
galaxies at optical wavelengths, and will be described elsewhere
\citetext{Noordermeer et al., in prep.}. \newline  
{\bf Upper middle:} Grayscale and contour representation of the velocity 
field. Dark shading indicates the receding side. Contours are spaced 
at 25 km~s$^{-1}$. The thick contour indicates the systemic
velocity from Table~\ref{table:data}. \newpage
{\bf Lower middle:} Position-velocity diagram along the major 
axis. Contours and grayscales show a slice through the data cube, the
squares show the corresponding velocities from the velocity field,
determined by fitting skewed Gaussians to the line profiles
(cf.\ Sect.~\ref{subsec:vfields}). 
Note that these are {\em not} rotation curves, and that in some
cases there are clear and pronounced deviations from the projected
rotational velocities (e.g.\ UGC~2183; cf.\ Sect.~\ref{subsec:vfields} 
and individual notes in Sect.~\ref{sec:notes}).
Contours are at -1.5 and -3
(dotted) and $1.5,3,6,12,\ldots$ times the rms noise in the channel
maps. The  
position angle along which the slice is taken is indicated in the top
left, the systemic velocity from Table~\ref{table:data} and central
position are indicated with the dashed lines. \newline 
{\bf Upper right:} Global profile. The dashed line indicates the
systemic velocity from Table~\ref{table:data}. Note that we always
used the Hanning-smoothed data cube at $60\arcsec$ resolution to
produce the global profile (except for UGC~6621 and 6623, see
Sect.~\ref{sec:notes}). \newline 
{\bf Lower right:} The radial \HI\ surface density profile. The dotted line
indicates the profile for the receding half of the galaxy, the dashed
line the approaching side. The thick solid line gives the profile
averaged over the entire disk. Data are only shown when above the
2$\sigma_{\mathrm {map}}$-level in the \HI-map. Profiles with symbols
are determined by measuring the intensity on concentric ellipses;
profiles without symbols are for edge-on galaxies and are determined
using Warmels's method (see Sect.~\ref{subsec:radprofs} for
details). \newline  

\clearpage

\begin{figure*}[tb]
  \centerline{\psfig{figure=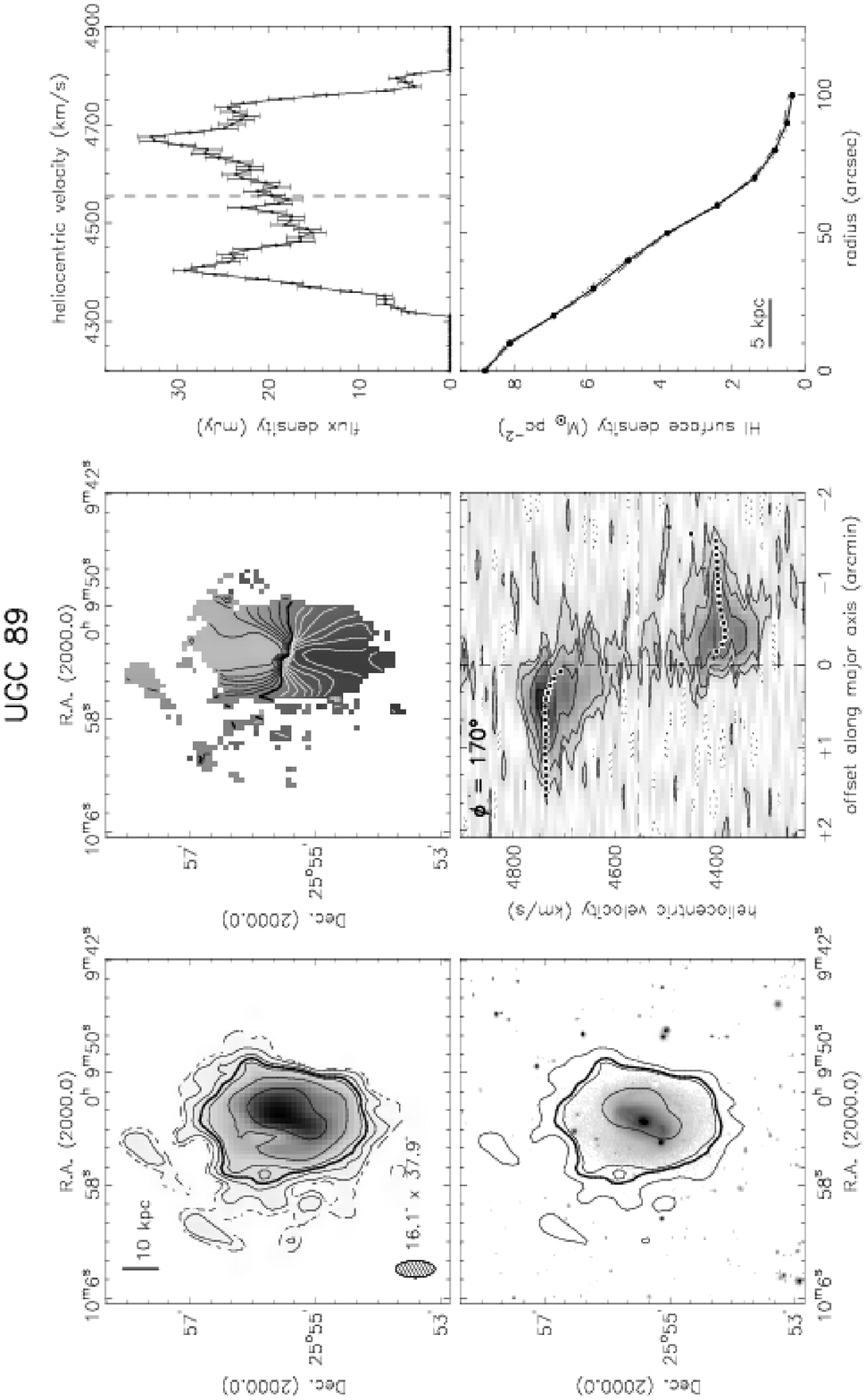,height=4.6in,angle=-90}}
\end{figure*}
\begin{figure*}[tb]
  \centerline{\psfig{figure=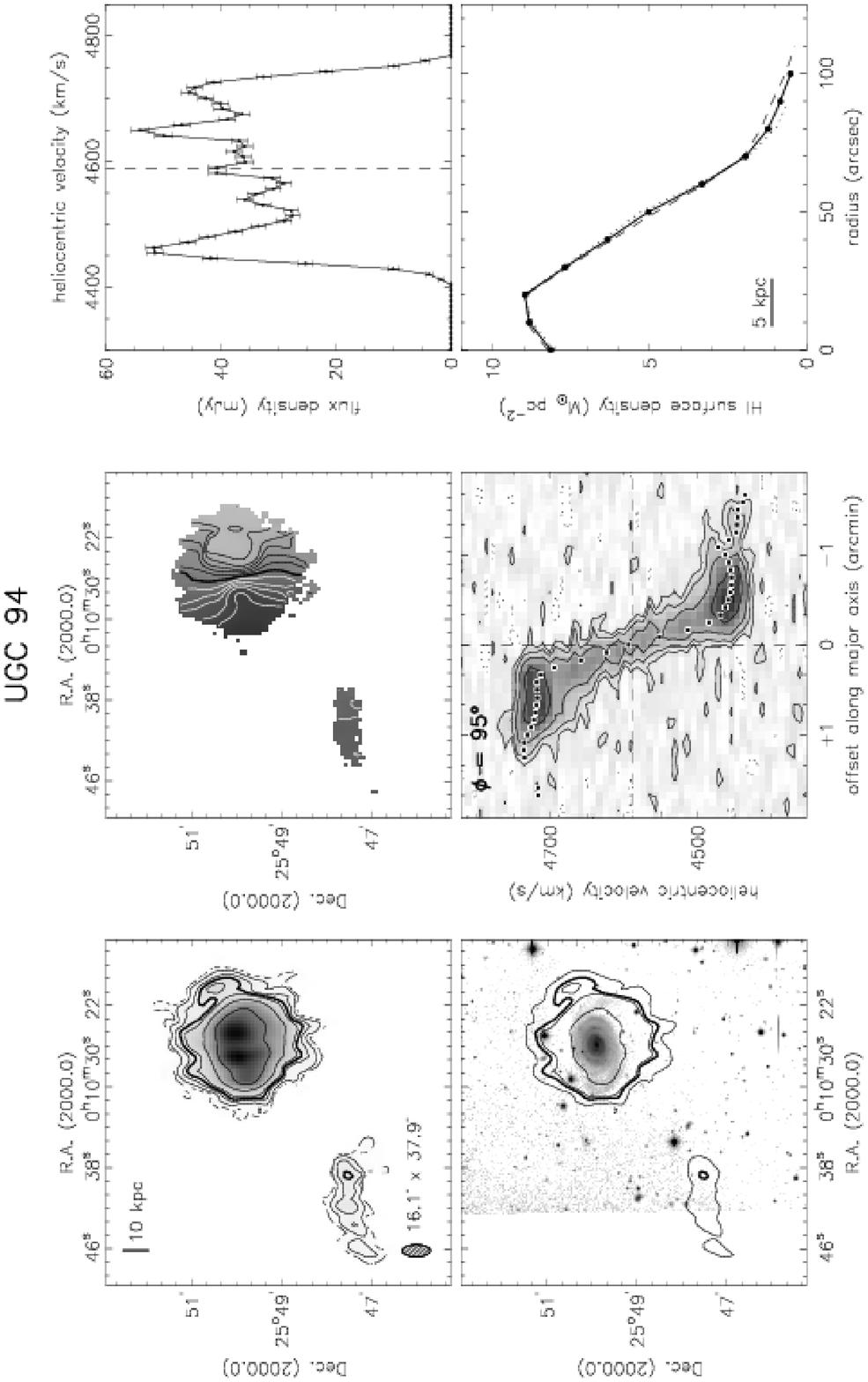,height=4.6in,angle=-90}}
\end{figure*}
\clearpage

\end{document}